\documentclass[11pt,titlepage]{article}
\usepackage[margin=1.2in]{geometry}
\usepackage{amsmath, amsbsy, amsthm, array, mathrsfs}
\usepackage{graphics}
\usepackage{physics}
\usepackage{epsfig, amssymb,latexsym,verbatim}
\usepackage{graphicx}
\usepackage{color}
\usepackage{dsfont, bbm}
\usepackage{relsize}

\newcommand{\R}{\mathbb{R}}
\newcommand{\mbs}{\mathbb{S}}

\newcommand{\noin}{\noindent}
\newcommand{\bee}{\begin{eqnarray*}}
\newcommand{\ene}{\end{eqnarray*}}
\newcommand{\bec}{\begin{center}}
\newcommand{\enc}{\end{center}}
\newcommand{\be}{\begin{equation}}
\newcommand{\ee}{\end{equation}}

\newcommand{\mb}{\mathbf}
\newcommand{\bs}{\boldsymbol}
\newcommand{\tb}{\textbf}
\newcommand{\pend}{$\blacksquare$}
\newcommand{\vs}{\vskip 3mm}
\newcommand{\bi}{\begin{itemize}}
\newcommand{\ei}{\end{itemize}}
\begin{document}

\title{\LARGE  On general notions of depth for regression
 \\[4ex]
}

\author{ {\sc Yijun Zuo}\\[2ex]
         {\small {Yijun Zuo  is professor at  Department of Statistics and Probability} }\\
         {\small Michigan State University, East Lansing, MI 48824, USA} \\
         {\small zuo@msu.edu }\\[6ex]
     }
\date{\today}
\maketitle

\vskip 3mm
{\small

\begin{abstract}
Depth notions in location have generated tremendous attention in the literature. In fact, data depth and its applications remain as one of the most active research topics in statistics over the last three decades.
Most favored notions of depth in location include Tukey (1975) halfspace depth (HD), Liu (1990) simplicial depth,
and projection depth (PD) (Stahel (1981) and Donoho (1982), Liu (1992), Zuo and Serfling (2000) (ZS00) and Zuo (2003)), among others.
\vskip 3mm

Depth notions in regression have also been proposed sporadically, nevertheless. The regression depth (RD) of Rousseeuw and Hubert (1999) (RH99), the most famous, exemplifies a direct extension of Tukey HD to regression.
Other notions include Carrizosa (1996) and the ones proposed in this article via modifying a functional in Marrona and Yohai (1993) (MY93).
 Is there any relationship between Carrizosa depth and the RD of RH99?
 Do these depth notions possess desirable properties? What are the desirable properties?   Can existing notions serve well as depth notions in regression? These questions remain open.
  \vskip 3mm
 The major objectives of the article include
(i) revealing the connection between Carrizosa depth and RD of RH99; (ii)
expanding location depth evaluating criteria in ZS00 for regression depth notions; (iii) examining the existing regression notions with respect to the gauges; and (iv) proposing the regression counterpart of the eminent location projection depth.

\bigskip
\par

\noindent{\bf Key words and phrase:} Depth, unfitness, linear regression, maximum depth regression estimating functionals, robustness.
\bigskip
\par
\noindent {\bf Running title:} Depth notions in regression.
\end{abstract}
}

\section{Introduction}
Notion of depth in location has attracted vast attention and has been increasingly pursued as a powerful tool for multi-dimensional nonparametric  data analysis and inference.\vskip 3mm

Prevailing location depth notions include Tukey (1975) halfspace depth (HD)(popularized by Donoho and Gasko (1992)),  Liu (1990) simplicial depth, the spatial depth (Vardi and Zhang (2000)), projection depth (PD) (Stahel (1981) and Donoho (1982), Liu (1992), Zuo and Serfling (2000) (ZS00), and Zuo (2003)), and zonoid depth (Koshevoy and Mosler (1997), Mosler (2002)).  Applications of data depth in multivariate statistics include
\begin{itemize}
\item[] (i) construction of multivariate inference procedures, such as depth-based tests, rank tests, multivariate quantiles, control charts, and confidence regions (Liu (1992), Liu (1995), Liu and Singh (1993), Liu, et al (1999),  Li and Liu (2004), Hallin et al (2010), Chakraborty and Chaudhuri (2014), Chernozhukov, et al (2017), Yeh and Singh (1997), Zuo (2009, 2010));
\item[] (ii) multivariate exploratory data analysis, such as in geophysical, hydrological, and physio meteorological research (Liu, et al (1999), Chebana and Ouarda (2008, 2011));
\item[] (iii) outlier detection, such as in environmental studies (Dang and Serfling (2010), Serfling and Wang (2014), Wang and Serfling (2015, 2018), Febrero, et al (2007));
\item[](iv) clustering and classification, especially in microarray gene expression data analysis, discriminant analysis, and supervised learning (Hoberg (2000), J$\ddot{o}$rnsten (2004), Ghosh and Chaudhuri (2005), Mosler and Hoberg (2006),  Cui, Lin, and Yang (2008),  Li, et al (2012), Lange, Mosler and Mozharovskyi (2014),  Paindaveine and Van Bever (2015));
\item[] (v)  multivariate risk measurement in financial engineering (Cascos and Molchanov (2007));
\item[] (vi) remote sensing and signal processing (Velasco-Forero and Angulo (2011, 2012));   robust linear programming (Mosler and Bazovkin (2014)), and
\item[] (vii) econometric and social studies (Caplin and Nalebuff (1988, 1991a, 1991b)),
\end{itemize}
 among others. In fact, data depth and its applications remain as one of the most active research topics in statistics over the last three decades.
 \vskip 3mm

The notion of location depth has been extended to local depth (Agostinelli and Romanazzi (2011), Paindaveine and van Bever (2013)); to depth for concentration, scatter and shape matrices (Chen, et al (2015), Paindaveine and van Bever (2017)); and to depth for functional data (e.g., L\'{o}pez-Pintado and Romo (2009), Claeskens, et al (2014), Hubert, et al (2015), Nieto-Reyes and Battey (2016), Gijbels and Nagy (2017), among others).
\vskip 3mm
Mizera (2002) introduced a scheme as a calculus technique/tool to derive depth functions for different statistical models, and Mizera and M\"{u}ller (2004) extended the tangent depth for location and regression to the location-scale setting. Data depth has also been employed as penalties in penalized regression (Majumdar and Chatterjee (2017)).
\vskip 3mm

Depth notions in regression have been inevitably proposed, yet sporadically. Regression depth ($\text{RD}$) by Rousseeuw and Hubert (1999) (RH99), the most famous, exemplifies a direct extension of Tukey location depth in regression. Others include pioneer Carrizosa depth (Carrizosa 1996) and the ones proposed here which are induced from Marrona and Yohai (1993) (MY93). Attention paid to regression depth has been disproportionally light, compared with its location counterpart. One of the reasons for this might be that there exist no clear fundamental principles to evaluate or measure proposed regression depth notions. Lack of the evaluation criteria not only prohibits any further advance and theoretical development of the depth notions in regression but also impedes their applications in practice. One major objective of the current article is to extend the set of criteria (or desired axiomatic properties) for depth notions in location in ZS00 to regression and to examine the existing regression notions of depth with respect to the proposed criteria.
\vskip 3mm

There exist a variety of robust methods including M-estimate approach and ad hoc ones (see Rousseeuw and Leroy (1987) (RL87), Maronna, Martin,and Yohai (2006) (MMY06)) for estimating the parameters in a linear regression model . In this article, regression depth is utilized to introduce the median-type deepest estimating functionals for regression parameters, manifesting one of the prominent advantages of notions of depth.
The functionals are the minimizers of the maximum of unfitness of regression parameters and recover in the empirical case the classical least squares, least absolute deviations, and other existing leading estimators.  Under a general framework, depth notions induced from projection-pursuit approach include the $\text{RD}_{RH}$ and projection regression depth (PRD) (induced from MY93) as special cases. The latter is an extension of the eminent PD in location to regression.
\vskip 3mm

The rest of this article is organized as follows. Section 2 presents a general definition for notions of unfitness and depth in regression and puts forward four general approaches for introducing the notions of unfitness or depth and the maximum (deepest) depth functionals while examining three special examples. It is found that Carrizosa (1996) depth, $D_C$, known of recovering the HD in location, is not identical but closed related to $\text{RD}_{RH}$ in regression. Section 3 provides a rigorous definition of depth (or unfitness) notion in regression based on four  axiomatic properties which then are employed for the evaluation of  four types of special depth notions. Section 4 ends the article with brief concluding remarks.
 Section 5 (Appendix) collects some major proofs and derivations and auxiliary lemmas. \vskip 3mm
\section{Definitions, approaches, and examples}

\subsection{Regression model} 
Consider a general linear regression model:
\begin{eqnarray}
y&=&\mathbf{x}'\boldsymbol{\beta}+{{e}}, \label{eqn.model}
\end{eqnarray}
where $'$ denotes the transpose of a vector;  random variable $y$ and ${e}$ are in $\R^1$;  and random vector $\mathbf{x}=(x_1,\cdots, x_p)'$ and  parameter vector $\boldsymbol{\beta}$ 
are in $\R^p$. 
Note that this general model includes the special case with an intercept term. For example,
if $\bs{\beta}=(\beta_1, \bs{\beta_2}')'$ and $x_1=1$, then one has $y=\beta_1+\mb{x_2}'\bs{\beta_2}+{e}$, where $\mb{x_2}=(x_2,\cdots, x_p)' \in \R^{p-1}$.
Denote $\mb{w}=(1,\mb{x_2}')'$, then $y=\mb{w}'\bs{\beta}+{e}$. We use this model or (\ref{eqn.model}) interchangably, depending on the context. Denote by $F_{(y,~\mathbf{x})}$ the joint probability distribution of  $y$ and $\mathbf{x}$ under the model (\ref{eqn.model}).
\vskip 3mm

In the following sections, we discuss the notions of \emph{unfitness} or \emph{depth} and \emph{general approaches} to introduce regression depth and induced deepest estimating functionals.\vskip 3mm

\subsection{Notions of unfitness and depth in regression} 
\label{unfitness.depth.sec}
\emph{Unfitness} of a candidate  parameter $\boldsymbol{\beta}$: UF$(\boldsymbol{\beta})$, is a function of the residual
$r(\boldsymbol{\beta}):=(y- \mathbf{x}'\boldsymbol{\beta})$.
Namely, UF$(\boldsymbol{\beta})=f(r(\boldsymbol{\beta}))$.
Examples of $f(x)$ include,
$x^2$ and $|{x}|$.
  Generally speaking, an even, monotonic in $|x|$, and convex $f (\cdot)$ with its minimum value $0$ at $0$ will serve the purpose.
  \vskip 3mm

 \emph{Depth} of $\bs{\beta}$ then can be defined as a bounded reciprocal (reverse) function of $\phi(F_R)$ (e.g. $1/(1+x)$), say on $[0,1]$, where $\phi$ is a functional on the distribution of $R:=\mbox{UF}(\bs{\beta})$. A typical example of $\phi$ is the expectation or quantile functional, $\phi(F_R)$ could also just be $R$.  Likewise, given its depth, one can define the unfitness of $\bs{\beta}$ to be a reciprocal function of the depth. 
 \vskip 3mm
A minimizer $\bs{\beta^*}$ of unfitness function  of $\bs{\beta}$ over all $\bs{\beta}\in\R^p$ can serve as a regression estimating functional for $\bs{\beta}$.
 Similarly, a maximizer of depth function plays the same role. 
\vskip 3mm
\subsection{Four approaches for notions of unfitness and depth}
\subsubsection{Classical objective function approach} \label{classical-sec}
 Directly employing the scheme above, one can recover many classical regression estimators in the empirical distribution case (i.e. $r_i(\bs{\beta})=y_i-\mb{x_i'}\bs{\beta}$, $i=1,\cdots, n$). Here the classical objective function in regression serves as the unfitness function, that is: {UF}$(\bs{\beta})=f_{Obj}(r(\bs{\beta}))$.
Maximizing depth of $\bs{\beta}$ is  equivalent to the minimization of $\phi( F_{R})$ and then the minimizer denoted by $\bs{\beta}^*$ could serve as an \emph{estimator} for $\boldsymbol{\beta}\in \R^p$.
In the sequel, consider examples of $\phi$: \tb{(i)} the expectation functional $\bs{\mu}$, and \tb{(ii)} quantile functional $\bs{q}_\tau$, $\tau\in (0,1)$.
\vskip 3mm

\noindent
\textbf{Example 2.1}\vskip 3mm

(I) If $\phi=\bs{\mu}$ and $f(x)=x^2$, then $\bs{\beta}^*(F_{(y,\mb{x})})=\arg\min_{\bs{\beta}\in R^p}\int(t-\mb{s}'\bs{\beta})^2dF_{(y,\mb{x})}(t,\mb{s})$, which  induces \emph{the least squares} (LS) estimator when $F_{(y,\mb{x})}$ is the empirical distribution.
\vskip 3mm
(II) If $\phi=\bs{\mu}$ and $f(x)=|x|$, then the approach above leads to \emph{the least absolute deviations} (LAD) estimator;
\vskip 3mm
(III) If $\phi=\bs{\mu}$ and  $f_\tau(x)=x(\tau-\mb{I}(x<0))$, $\tau \in (0,1)$, where $\mb{I}$ is the indicator function, then the approach above results in \emph{the quantile regression} estimator (Koenker and Bassett (1978)). When $\tau=1/2$ it recovers the $L_1$ regression estimator,  for related discussions on quantile regression, see Portnoy (2003, 2012);
\vskip 3mm
(IV) If $\phi=\bs{q}_{0.5}$ and $f(x)=x^2$, then the approach above yields \emph{the least median squares} (LMS) estimator (Rousseeuw (1984)); and
\vskip 3mm
(V) If $\phi=\bs{\mu}$, coupled with
an appropriately chosen function $f(x)$, one  can actually recover the  M-estimators (Huber (1973) (including
the famous (a) Huber's proposal 2 (Huber (1964)), (b) Hampel's three-parts (Hampel (1974)), and (c) Tukey bisquare (Beaton and Tukey, (1974)) ones), the L-estimators (Ruppert and Carroll (1980)), and the R-estimators (Koul (1970, 1971), Jureckova (1971), and Jaeckel (1972)). \hfill \pend
\vskip 3mm
\subsubsection{Facility location approach} \label{carrizosa.sec}
In addition to the general approach mentioned in the Section \ref{classical-sec}, there exist other approaches for the introduction of notions of depth or unfitness. The classical one is the \emph{facility location approach}, prevailing in location analysis and operations research.\vskip 3mm

Let $\mb{F}_1 \in\R^2$ be a candidate for a facility location, and $P$ be the probability distribution of a random vector $\mb{X} \in \R^2$ (or of consumers' locations), and $d(\mb{F}_1,\mb{X})$ (defined in Remarks 2.1 below) measures in some sense the closeness of $\mb{F}_1$ to the distribution $P$ of $\mb{X}$ (or the coverage of consumers). Let $\mb{F}_2$ be a candidate for the facility location of  another competitive company. Similarly, $d(\mb{F}_2,\mb{X})$ measures the coverage of the  consumers in the vicinity of the facility at $\mb{F}_2$. With respect to the $\mb{F}_1$, the maximum market share which can be captured by any other facility is $\sup_{\mb{F}_2\in \R^2}P\left(\omega: d(\mb{F}_2, \mb{X(\omega)})<d(\mb{F}_1,\mb{X(\omega)})\right)$. Also, the $\mb{F}_1$ should be chosen to maximize its market share. 
\begin{align}
\mb{F}^*_1&=\arg\!\max_{\mb{F}_1 \in \R^2}\bigg( 1-\sup_{\mb{F}_2\in \R^2}P\left(\omega: d(\mb{F}_2, \mb{X(\omega)})<d(\mb{F}_1,\mb{X(\omega)})\right)\bigg) \nonumber \label{x.eqn}  \\[1ex]
&= \arg\!\max_{\mb{F}_1 \in \R^2}\inf_{\mb{F}_2\in \R^2 }P\left(\omega: d(\mb{F}_1, \mb{X(\omega)})\leq d(\mb{F}_2,\mb{X(\omega)})\right).
\end{align}
 Carrizosa (1996) extended $\R^2$ above to $\R^p$ ($p\geq 2$) and introduced a depth notion (normalized depth, see Definition 2.1 below). Let us use a generic term, \emph{Carrizosa depth}, hereafter. The Carrizosa depth of $\mb{x}$ w.r.t. $P$: $D_{C}(\mb{x}; P)$ ($P$ and $F_{\mb{X}}$ are used interchangeably),  is defined as
 \be
 D_{C}(\mb{x};P)=\inf_{\mb{y}\in \R^p}P\left(\omega:d(\mb{x}, X(\omega))\leq d(\mb{y},X(\omega))\right).\label{C-D.eqn}
 \ee
 Then $\mb{F}^*_1$ in (\ref{x.eqn}) is the maximum depth solution (functional) for the facility location problem. \vskip 3mm

 \noindent
 \tb{Remarks 2.1}\vskip 3mm
 (I)
 Note that the distance measure $d$ above includes a class of possible choices. For example, $d$ could be an
 $L_p$ norm or weighted $L_p$ norm (see Zuo (2004)) ($p\geq 1$). \vskip 3mm

 (II) When $d (x,y)=\|x-y\|$, where ``$\|\cdot\|$'' stands for the Euclidean (or $L_2$) norm,
 (\ref{C-D.eqn}) recovers the\emph{ normalized depth} $\mbox{ND}(x; P)$ of Carrizosa (1996) that is quoted below:\vs
 \noin
 \tb{Definition 2.1} (Carrizosa 1996). The normalized depth of ND$(x; P)$ of a point $\mb{x}\in \R^p$ in $P$ is defined as
 \be \mbox{ND}(\mb{x}; P)=\inf_{\mb{y}\in \R^p}P(\{a:\|y-a\|\geq \|x-a\|\}) \label{nd.eqn}
  \ee
  Henceforth we focus on $L_2$ norm for distance measure $d$, {unless stated otherwise}. \hfill \pend
  \vskip 3mm

 In the location setting, Donoho and Gasko (1992) first addressed the notion of depth proposed in Tukey (1975),
 their empirical depth is some integer among $\{1,\cdots, n\}$.
 In the following, we invoke a slightly different characterization  for Tukey's depth given in Zuo (1998)(called halfspace depth (HD)).
 \be
 \mbox{HD}({x};P)=\inf_{H}\big\{P(H):~
 \text{ $H$ is a closed halfspace and ${x}\in H$ }
 \big\}, ~ x\in \R^p. \label{hd.eqn}
 \ee

 It turns out that $D_C(x;P)$ can actually recover $\mbox{HD}(x;P)$ as shown in Carrizosa (1996).\vskip 3mm

 \noindent
 \tb{Proposition 2.1} If  $d$ in (\ref{C-D.eqn}) is the $L_2$  norm, then
 $D_C(x;P)$, equivalently $\mbox{ND}(\mb{x}; P)$ in (\ref{nd.eqn}), is identical to $\mbox{HD}(x;P)$ in (\ref{hd.eqn}). \vskip 3mm

\noin
{\sc Proof}: see the proof of Proposition 1 of  Carrizosa (1996). 
\hfill \pend

\vs
 Although the depth $D_C(x;P)$ above is introduced initially for the location problem, it can be 
 extended for the regression problem, as done in Carrizosa (1996) with the $L_1$ norm.\vskip 3mm

Indeed, given a probability measure $P$ (or equivalently $F_{(y,\mb{x})}$) in $\R^{p}$,
one could define the
$D_C(\bs{\beta}; P)$ of  $\bs{\beta} =(\beta_1,\bs{\beta_2}')'
\in\R^{p}$ as follows: for $|\beta_1|< \infty$
\be
D_C(\bs{\beta}; P)= \inf_{\bs{\alpha} \in \R^{p}}
P\left(~d(y, (1,\mb{x}')'\bs{\beta})\leq d(y, (1,\mb{x}')'\bs{\alpha})~\right), ~\mb{x} \in \R^{p-1}, (p\geq 2)
 \label{D-C-regssion.eqn}
\ee
where 
$\bs{\alpha}=(\alpha_1, \bs{\alpha}_2')'$ and  $\bs{\alpha}_2, ~\bs{\beta_2} \in \R^{p-1}$;
if $|\beta_1|\to \infty$, then define $D_C(\bs{\beta};P)\to 0$.
When 
 $d(x,y)=|x-y|$, (\ref{D-C-regssion.eqn}) recovers the \emph{depth in regression} in Carrizosa (1996). 
The latter seems to be the pioneer notion of  depth in regression  in the literature. Does it have anything to do with
 the 
  $\mbox{RD}_{RH}$ of RH99? Let us first quote the original definition of RH99.\vs
 \noin
\tb{Definition 2.2} (RH99). The regression depth of $\bs{\beta}$ is the minimum probability mass that needs to be passed when tilting $\bs{\beta}$ in any way until it is vertical.
 \vs
 Since $D_C$ recovers HD in location and $\mbox{RD}_{{RH}}$ is an extension of HD in regression,  naturally, one wonders whether $D_C$ can recover $\mbox{RD}_{{RH}}$ in regression. The two are closely connected but not identical as revealed in Proposition 2.2  below.
\vskip 3mm
The same idea of Carrizosa (1996) was  proposed again in Adrover, Maronna, and Yohai (2002) (AMY02). AMY02 first flawlessly defined $\mbox{RD}_{{RH}}$ above to be
\be
\mbox{RD}_{RH}(\bs{\beta}, P)=\inf_{\bs{\lambda}\neq \mb{0}}P\left( \frac{r(\bs{\beta})}{\bs{\lambda}'\mb{x}}<0, \bs{\lambda}'\mb{x}\neq 0\right), \label{amy02-d1.eqn}
\ee
under assumptions (a) and (b) below,
where, $\bs{\lambda}\in \R^p$,  $r(\bs{\beta})=y-\mb{x}'\bs{\beta}$.  They then proposed the depth: 
\be
D(\bs{\beta}, P)=\inf_{\bs{\gamma} \in \R^p}P(|r(\bs{\beta})|\leq |r(\bs{\gamma})|). \label{amy02-d2.eqn}
\ee
 If the first coordinate $x_1$ of $\mb{x}$ in (\ref{amy02-d2.eqn}) is $1$ and $d(x,y)=|x-y|$ in (\ref{D-C-regssion.eqn}), then (\ref{amy02-d2.eqn}) and (\ref{D-C-regssion.eqn}) coincides.
\vs
Under the assumptions  (a) $P(\mb{x}'\mb{v}=0) =0$ for all $\mb{v}\neq \mb{0} \in \R^p$ and (b) $P(r(\bs{\beta})=0)=0$  for all $\bs{\beta} \in \R^p$, AMY02 showed that (\ref{amy02-d2.eqn})
is equivalent to (\ref{amy02-d1.eqn}) (the last step of the proof is debatable though).
(a) and (b) exclude \emph{any} discrete distribution cases of $(y,\mb{x})$, nevertheless.
 The following result characterizes $D_C(\bs{\beta};P)$ and reveals its connection with $\mbox{RD}_{RH}(\bs{\beta};P)$.
\vskip 3mm

Write $\mb{w}=(1,\mb{x}')'$ and $r(\bs{\beta})=y-\mb{w}'\bs{\beta}$. If $d(x,y)=|x-y|$, then (\ref{D-C-regssion.eqn})
is equivalent to
\be
D_C(\bs{\beta}, P)=\inf_{\bs{\alpha} \in \R^p}P(|r(\bs{\beta})|\leq |r(\bs{\alpha})|), \label{D-C.eqn}
\ee
For a given $\bs{\beta} \in\R^p$ with $\|\bs{\beta}\|<\infty$, denote by $H_{\bs{\beta}}$ the unique hyperplane determined by $y=\mb{w'}\bs{\beta}$.
Likewise, a given non vertical hyperplane $H$ 
uniquely identifies an $\bs{\alpha} \in \R^p$ through $y=\mb{w'}\bs{\alpha}$. Define $S(\bs{\beta}):=\{\bs{\alpha}\in\R^p: ~ H_{\bs{\alpha}} \mbox{~intersects with~} H_{\bs{\beta}}\}$ for the given $\bs{\beta}$.  \vs
\noin
\tb{Proposition 2.2} If $d(x,y)=|x-y|$ in (\ref{D-C-regssion.eqn}), then
(i) $D_C(\bs{\beta}; P)=P(r(\bs{\beta})=0)$, 
and (ii) $\mbox{RD}_{RH}(\bs{\beta};P)=\inf_{\bs{\alpha} \in S(\bs{\beta})}P\left(|r(\bs{\beta})|\leq |r(\bs{\alpha})|\right)$.
\vskip 3mm
 \noindent
{\sc Proof}: see the Appendix. \hfill \pend
\vs

$D_C(\bs{\beta};P)$ in (\ref{D-C.eqn}) is not identical to original $\mbox{RD}_{RH}$ of RH99, but is closely related to the latter. In fact, if the
infimum in RHS of (\ref{D-C.eqn}) performs over $S(\bs{\beta})$, then they are identical. This is another characterization of RD$_{RH}$. $D_C(\bs{\beta};P)$ is no greater than $\mbox{RD}_{RH}(\bs{\beta};P)$.\vs

Based on the depth functional in (\ref{D-C-regssion.eqn}), we can introduce
the maximum regression depth estimating functional for $\bs{\beta}$, which is defined, for $d(x,y)=|x-y|$, as
\be
\bs{\beta}^*(P)=\arg\!\!\max_{\bs{\beta} \in\R^{p}}D_C(\bs{\beta}; P), \label{D-C-beta^*.eqn}
\ee
$\bs{\beta}^*(P)$ above is well defined. That is, the maximum on the RHS of (\ref{D-C-beta^*.eqn}) is attained at a bounded $\bs{\beta}$. The latter  is safeguarded by the result below under the assumption:\\[1ex]
\hspace*{20mm}\tb{(A)}: ~~ $P(H_v)=0$ for any vertical hyperplane $H_v$
\vskip 3mm
\noin
\tb{Proposition 2.3} Under \tb{(A)}, 
  (i) $\lim_{\|\bs{\beta}\|\to \infty}D_C(\bs{\beta};P)=0$, and
(ii) the maximum on the RHS of (\ref{D-C-beta^*.eqn}) exists and is attained at a bounded $\bs{\beta}$.
\vskip 3mm
\noindent
{\sc Proof}: see the Appendix. \hfill \pend
\vs

\subsubsection{Projection-pursuit approach}

There is another approach based on the projection-pursuit (PP) scheme to induce the regression estimating functional for parameter $\boldsymbol{\beta}$. One starts with
a univariate regression estimating functional w.r.t.  the univariate variable $\mb{u}'\mb{x} \in \R$ and $r(\bs{\beta})$ along each direction $\mb{u}\in \mbs^{p-1}:=\{\mb{v},\|\mb{v}\|=1, \mb{v}\in\R^p\}$
and calculates the $\mbox{UF}_{\mb{u}}(\boldsymbol{\beta})$ (the unfitness along $\mb{u}$)(see Section \ref{unfitness.depth.sec} for the definition of unfitness). Then one obtains $\text{UF}(\bs{\beta})$, the supremum of  $\mbox{UF}_{\mb{u}}(\boldsymbol{\beta})$ over all $\mb{u}\in \mbs^{p-1}$. Finally, one minimizes $\text{UF}(\bs{\beta})$ over all $\bs{\beta}\in\R^p$ 
to obtain a regression estimating functional $\bs{\beta}^*$ for $\boldsymbol{\beta}$ via the min-max scheme.
\vskip 3mm

\noindent
\textbf{Remarks 2.4}\vskip 3mm
(I) The approach above actually can 
 recover the maximum regression depth functional in RH99 and induce a maximum projection depth functional that is closely related to P1-estimate in MY93. We elaborate the two special cases in the following.\vskip 3mm
(II) A  related PP approach was discussed in RL87 (page 144). In the empirical case, to obtain the estimator $\bs{\beta}^*$, it minimizes an objective dispersion function $s(r_1(\bs{\beta}),\cdots, r_n(\bs{\beta}))$, where $r_i(\bs{\beta})=y_i-\mb{x}'_i\bs{\beta}$, $s$ is just scale equivariant (not translation invariant), and $r_i(\bs{\beta})$ are regarded as
a projection of the point $(y_i,\mb{x}'_i)'$ onto $(1,-\bs{\beta})'$. By varying $s$, this approach covers a very large family  of estimators (LS, LAD, LTS, LMS, and S, etc.).
\hfill\pend
\vskip 3mm
\noindent
\textbf{Example 2.2. Regression depth and Maximum regression depth functional} \label{rd.exam}\\[1ex]
\noindent
Consider the linear model: $y=\beta_1+\mathbf{x}'\boldsymbol{\beta}_2+{e}$, where $\mathbf{x},\boldsymbol{\beta}_2 \in \R^{p-1}$.
Denote
$\mathbf{w}=(1,\mathbf{x}')'$, $\boldsymbol{\beta}=(\beta_1, \boldsymbol{\beta}_2')'$. Then
the model is: $y= \mb{w}'\bs{\beta}+{e}$. That is, $\mb{w}$ here corresponds to $\mb{x}$ in general model (\ref{eqn.model}) and vice versa.\vskip 3mm

When $p=2$, we define $\mbox{F}(\boldsymbol{\beta})=E(\mathbf{I}\left((y-\mathbf{w}'\boldsymbol{\beta})*\mb{v}'\mathbf{w}\geq 0\right))$, where $F(\boldsymbol{\beta})$ stands for ``fitness'' of $\bs{\beta}$, $\mathbf{I}$ for the indicator function,
 and $\mb{v}=(-v_1,{v}_2)$, $v_1\in\R$, $|{v}_2|=1$. When, $v_2=1$, it represents the total probability mass touched (covered) by tilting   the line
 $y={\beta_1}+{\beta_2}x$ \emph{counter-clockwise} around the point $(v_1, \beta_1+\beta_2 v_1)$  to the vertical position (note that the point is the intersection point of the line with the vertical line $x=v_1$). By considering  the clockwise tilting ($v_2=-1$), it is seen that the closer to $1/2$ the total mass is, the better (more balanced) the candidate parameter $\bs{\beta}$ is.
 \vskip 3mm
When $p>2$, with the same $ \mbox{F}(\boldsymbol{\beta})$ as defined above,
it can be shown that in the empirical case, minimizing $\mbox{F}(\boldsymbol{\beta})$ over all $v_1\in \R$ and $\mb{v}_2$ with $\|\mb{v}_2\|=1$  ($\mb{v}=(-v_1,\mb{v}'_2)'$) 
 leads essentially to the regression depth $\mbox{RD}_{RH}$ of $\boldsymbol{\beta}$ in RH99
(See the derivations in the Appendix for the general case where it is shown that the approach here is equivalent to (\ref{RS04.eqn}) below). That is,
\be
\inf_{v_1\in \R, \|\mb{v}_2\|=1}E(\mathbf{I}\left((y-\mathbf{w}'\boldsymbol{\beta})*\mb{v}'\mathbf{w}\geq 0\right))=\mbox{RD}_{RH}(\bs{\beta}; P).
\ee

\vskip 3mm

Equivalent definitions (or characterizations) of RD$_{RH}$ in Definition 2.2 exist in the literature (see Remarks 5.1).
 The one given   
 in Rousseeuw and Struyf (2004) (RS04) is
\be
\mbox{RD}_{RH}(\bs{\beta}; P)= \inf_{D\in{\mathcal{D}}}\left\{P\left((r(\bs{\beta})\geq 0) \cap D \right)+P\left((r(\bs{\beta})\leq 0) \cap D^c \right)\right\},\label{RS04.eqn}
\ee
where $\mathcal D$ is the set of all vertical closed halfspaces D. 

Now, we can define  $\mbox{UF}(\boldsymbol{\beta})$ as a simple reciprocal function of $\mbox{F}(\boldsymbol{\beta})$ (e.g. $f(x)=a(1-x)/x$, $a>0$) such that it equals  $\infty$ if the latter  equals zero and equals zero if the latter is $1$.
Maximizing $\mbox{UF}(\boldsymbol{\beta})$ leads to the $\mbox{RD}_{RH}$ of $\boldsymbol{\beta}$. 
Furthermore, minimizing the maximum of $\mbox{UF}(\boldsymbol{\beta})$ over all $\boldsymbol{\beta} \in \R^{p}$ leads to the maximum regression depth functional $\bs{\beta}^*$. 
\hfill \pend
\vs
\noindent
\textbf{Example 2.3. Projection regression depth and maximum depth functional }\\[1ex]
Hereafter,  assume that $R$ is a {univariate}
 regression estimating functional
 which satisfies\\[1ex]
\noindent
(\tb{A1}) {regression}, {scale} and {affine equivariant}, that is, \vskip 2mm

$R(F_{(y+x{b},~{x})})=R(F_{(y,~{x})})+{b}$, ~$\forall ~{b}\in \R$; ~~~ \vskip 2mm

$R(F_{(sy,~{x})})=sR(F_{(y,~{x})})$, ~$\forall ~s \in\R$;  ~~~and \vskip 2mm

 $R(F_{(y,~a{x})})=a^{-1}R(F_{(y,~{x})})$,~ $\forall ~a\in\R$ and $a\neq 0$. \vskip 2mm

respectively, 
where $x, y\in \R$ are random variables.\vskip 3mm

\noindent
(\tb{A2}) $\sup_{~\|\mb{v}\|=1}|R(F_{(y, ~\mathbf{x}'\mb{v}})|\leq \infty$. 
\vskip 3mm

\noindent
(\tb{A3}) $R(F_{(y-\mathbf{x}'\boldsymbol{\beta},~\mathbf{x}'\mb{v})})$ is quasi-convex and continuous in $\boldsymbol{\beta}\in \R^p$ for any fixed $\mb{v}\in\mbs^{p-1}$.
\vskip 5mm

\noindent
Assume that $S$ is a positive scale estimating functional such that
 $S(F_{sz+b})=|s|S({F_z})$ \text{for random variable $z \in \R$ and scalar $b, s \in \R$}; that is, $S$ is \emph{scale equivariant} and \emph{location invariant}. 
 \vskip 3mm

 \noindent
 \textbf{Remarks 2.5}\vskip 3mm

  (I) Note that, the $R$ above applies for  the regression models that do not contain an intercept term (regression through the origin). The latter situation is required in certain applications (see page 62 in RL87) or is generally applicable by some simple treatments of original data (see Eisenhauer (2003)).\vskip 3mm
  (II) In the sequel, $R$ will be restricted to 
  the form $R(F_{(y-\mb{x'}\bs{\beta}, ~\mb{x'}\mb{v})})=T\big(F_{(y-\mb{x}'\bs{\beta})/{\mb{x}'\mb{v}}}\big)$, $\mb{x}'\mb{v}\neq 0$. $T$ could be a univariate location functional that is location, scale and affine equivariant 
  (see pages 158-159 of RL87 for definitions). 
   Examples of $T$ include mean and quantile functionals, among others. An example of R is   
   $R(F_{(y-\mb{x'}\bs{\beta},~\mb{x}'\mb{v})})=\text{Med}\big( (y-\mb{x'}\bs{\beta})/\mb{x}'\mb{v}\big)$, where Med stands for the median functional and
   $\text{Med}(Z)$ for
  $\text{Med}(F_Z)$. 
 \vskip 3mm
   (III) (A2) holds trivially if $T$ is a quantile-type functional (such as median functional) or  mean-type functional if the moments of the underlying distribution exist. 
    (A3) holds  for those $T$  as long as integrands involved are quasi-convex and continuous. 
 \hfill \pend
 \vskip 3mm

 Pairs of $T$ and $S$ induce a class of projection regression estimating functionals.
  Define
\be\mbox{UF}_\mb{v}(\boldsymbol{\beta}; ~F_{(y,~\mathbf{x})}, T):= |T\big(F_{(y-\mb{x}'\bs{\beta})/{\mb{x}'\mb{v}}}\big)|/S(F_{y}), \label{eqn.uf}
\ee
which represents
 unfitness of $\boldsymbol{\beta}$ at $F_{(y,~\mathbf{x})}$  w.r.t. $T$  along the direction $\mb{v}\in \mbs^{p-1}$.
 Note that if $T$ is a Fisher consistent estimating functional, then $T\big(F_{(y-\mb{x}'\bs{\beta})/{\mb{x}'\mb{v}}}\big)=0$ 
 under the assumption $E({e}|\mathbf{x})=0$ in model (\ref{eqn.model}) for some  $\bs{\beta}_0$ (the true parameter of the model) and 
  the classical model assumption that $0$ is some kind of center of the error distribution, 
 and $\bs{x}$ and ${e}$ are independent. Note that $S(F_y)$ does not depend on $\mb{v}$ and $\bs{\beta}$. \vskip 3mm

 That is, overall one expects $|T|$ to be small and close to zero for a candidate $\bs{\beta}$, independent of the choice of $\mb{v}$ and $\mathbf{x}'\mb{v}$. \emph{The magnitude of $|T|$ measures the unfitness of $\bs{\beta}$ along the $\mb{v}$}.
 Dividing here by $S(F_{y})$  is simply to guarantee the scale invariance of $\mbox{UF}_\mb{v}(\boldsymbol{\beta}; ~F_{(y,~\mathbf{x})}, T)$. Taking the supremum over all $\mb{v}\in\mbs^{p-1}$ and suppressing $T$,  yields
 \be
 \mbox{UF}(\boldsymbol{\beta};~F_{(y,~\mathbf{x})})= \sup_{\|\mb{v}\|=1}\mbox{UF}_\mb{v}(\boldsymbol{\beta}; ~F_{(y,~\mathbf{x})}, T),\label{eqn.UF}\ee
 the\emph{ unfitness} of $\boldsymbol{\beta}$ at $F_{(y,~\mathbf{x})}$  w.r.t. 
 $T$. 
 Now applying the min-max scheme, we obtain the \emph{projection regression estimating functional}
 \be
 \bs{\beta}^*(F_{(y,~\mathbf{x})})=\arg\!\min_{\boldsymbol{\beta}\in \R^p}\mbox{UF}(\boldsymbol{\beta}; ~F_{(y,~\mathbf{x})}),\label{eqn.T*}
 \ee 
 \vskip 3mm
 \noindent
 \textbf{Remarks 2.6}\vskip 3mm
(I)~$\mbox{UF}(\boldsymbol{\beta};~F_{(y,~\mathbf{x})})$ corresponds to the \emph{outlyingness} $O(x, F_X)$, and  $\bs{\beta}^*(F_{(y,~\mathbf{x})})$ corresponds to the\emph{ projection median functional} $PM(F_X)$ in the location setting (see Zuo (2003)). In (\ref{eqn.uf})  (\ref{eqn.UF}) and (\ref{eqn.T*}), we have suppressed  $S$ since it does not involve $\mb{v}$ and is nominal (besides achieving the scale invariance).
$T$  in (\ref{eqn.UF}) and (\ref{eqn.T*}) is also suppressed for convenience.
\vskip 3mm

(II)
 A similar $\bs{\beta}^*$  was first studied in MY93, where it was called P1-estimate (denote it by $T_{P1}$, see (\ref{eqn.MY93})). However, they are  different. First,  MY93 did not talk about the ``unfitness" (or ``depth"). Second,
 the definition of $\bs{\beta}^*$ here is \emph{different from} $T_{P1}$ of MY93, the latter multiplies by $S(F_{\mb{v}'\mathbf{x}})$ instead of dividing by $S(F_y)$ in (\ref{eqn.uf}). 
 They instead defined the following
 $$
 A(\boldsymbol{\beta}, \mb{v}) =| R(F_{(y- \boldsymbol{\beta}'\mathbf{x}, ~\mb{v}'\mathbf{x})})| S(F_{\mb{v}'\mathbf{x}}),
 $$
 where $\mb{v}, \boldsymbol{\beta} \in \R^p$. Their P1-estimate is defined as
 \be
 T_{P1}=\arg \min_{\boldsymbol{\beta}\in \R^p}\sup_{\|\mb{v}\|=1}A(\boldsymbol{\beta}, \mb{v}).\label{eqn.MY93}
 \ee
 Later we will revisit $T_{P1}$ and explain why we divide by $S(F_y)$ in (\ref{eqn.uf}) instead of multiplying $S(F_{\mb{v}'\mathbf{x}})$.
 Note that $S(F_y)$ here could also be replaced by $S(F_{y-\boldsymbol{\beta}'\mb{x}})$.
 \vskip 3mm

  (III) The projection-pursuit idea here was first employed in a multivariate location setting by  Stahel (1981) and Donoho (1982) independently.
 \hfill \pend
  \vskip 3mm
\noindent
\tb{Projection regression depth (PRD)} ~
 One can also introduce the notion of projection depth in regression using
 the $\mbox{UF}(\boldsymbol{\beta}; ~F_{(y,~\mathbf{x})})$. For example, to make the depth 
 between $0$ and $1$, define a \emph{projection regression depth} (PRD) functional of $\boldsymbol{\beta}$ at $F_{(y,~\mb{x})}$ w.r.t. a pair $(T, S)$  as
 \be
 \text{PRD}\left(\boldsymbol{\beta}; ~F_{(y,~\mathbf{x})}\right)=\left(1+\mbox{UF}\big(\boldsymbol{\beta}; ~F_{(y,~\mathbf{x})}\big)\right)^{-1}.
 \label{eqn.PRD}
 \ee
It is readily seen that the LHS of (\ref{eqn.T*}) is also a\emph{ maximizer} of projection regression depth functional.
 For the specifical pair of $T$  and $S$ such as
 \bee
 T(F_{(y-\mb{x}'\bs{\beta})/ (\mb{x'}\mb{v})})&=&\text{Med}_{\mb{x'}\mb{v}\neq 0}\big\{\frac{y-\mb{x}'\bs{\beta}}{\mb{x'}\mb{v}}\big\}, ~~~ 
 S(F_y)~~= ~~\text{MAD}(F_y), \label{S.eqn}
 \ene
we have
 \be
 \text{UF}(\bs{\beta}; F_{(y,~\mb{x})})=\sup_{\|\mb{v}\|=1}\Big|\text{Med}_{\mb{x'}\mb{v}\neq 0}\big\{\frac{y-\mb{x}'\bs{\beta}}{\mb{x'}\mb{v}}\big\}\Big|\bigg/ \text{MAD}(F_y),
 \ee
 and
 \be
 \text{PRD}\left(\bs{\beta}; ~F_{(y,~\mathbf{x})}\right)=\inf_{\|\mb{v}\|=1,\mb{x'}\mb{v}\neq 0}
 \frac{\text{MAD}(F_y)}{\text{MAD}(F_y)+\Big|\text{Med}\big\{\frac{y-\mb{x}'\bs{\beta}}{\mb{x'}\mb{v}}\big\}\Big|}. \label{special-PRD.eqn}
 \ee\vskip 3mm

The empirical case of PRD above is closely related to the so-called ``centrality" in Hubert, Rousseeuw, and Van Aelst (2001) (HRVA01). In the definition of the latter, all the terms of ``MAD$(\cdot)$" on the RHS  of (\ref{special-PRD.eqn}) are divided by $\text{Med}|\mb{x'}\mb{v}|$.

 \vskip 3mm

 \subsubsection{Other approaches}

 Besides the three approaches above, there are certainly other approaches (including ad hoc ones). Among them, Mizera (2000) is a famous one. 
  \vs
 In extending the idea of $\mbox{RD}_{RH}$ of RH99, Mizera (2002) (M02), with a decision-theoretic flavor and under the
 vector optimization framework (vector differential approach), introduced the notions of global, local and tangent depth rigorously. 
 The former two are based on the
 so-called ``critical" function. The latter (the tangent depth) is based on the vector differential approach  and includes local depth as a special case. The local depth in turn includes the global depth as its special case. They are identical under certain conditions.\vskip 3mm

With mainly Euclidean norm (and/or $L_1$ norm) of $X-\theta$ (in location) and of $y-\mb{x'}\bs{\beta}$ (in regression) as \emph{the typical critical functions}, M02 applied the notions of depth to location
and (linear, nonlinear, and orthogonal)  regression models and obtained specific depth functions in those models \emph{recovering mainly} both the HD of Tukey (1975) in location and $\mbox{RD}_{RH}$ in linear regression under a single unified notion of depth (the tangent depth).
It is not difficult to see that the critical function could be regarded as a form of unfitness measure of the underlying parameter (note that the words ``unfitness", ``nonfit", ``critical function", ``objective function", and ``loss function"
are interrelated in some sense. Different people have different preferences.)  For the linear regression model, the critical function in M02 can be summarized as follows:
\be
\text{CF}(\bs{\beta}; F_{(y,~\bs{x})})=c_p\|y-\bs{x}'\bs{\beta}\|_p,\label{Mizera-cf.eqn}
\ee
where $\|\cdot\|_p$  is the absolute value or squared value w.r.t. $p=1$ or $2$ respectively, and $c_p=1/p$. The global depth of this leads to $\mbox{RD}_{RH}$ of RH99.
 \vskip 3mm

Based on the definitions of M02, one can introduce notions of depth in regression models with
appropriate chosen critical functions. The key issue is how to construct ``reasonable" or ``optimal" critical functions besides the $L_1$ norm and the $L_2$ norm approaches given in M02. 
With the depth functions obtained via M02 approach, one can introduce the maximum (deepest) regression depth estimating functionals via the min-max scheme.   \hfill \pend

 \vskip 3mm
\vs
In addition to the approaches we have discussed so far, there are certainly other ones which introduce notions of unfitness or depth. Can all these notions really serve as depth notions in regression?  Gauging or evaluating those notions naturally becomes an issue. Namely, all the unfitness or depth notions must satisfy some basic desired axiomatic properties or possess some desirable and intrinsic features and meet some criteria. What are the criteria? \vskip 3mm

 In the following, we will propose and discuss four axiomatic properties that are deemed necessary for any notion of regression depth or unfitness, thereby providing a systematic basis for the selection and evaluation of a depth notion in regression.
\section{Axiomatic properties for depth and unfitness}
\subsection{Four Axiomatic properties}
\vs
\tb{Definition 3.1} (A depth notion in regression) \vs  A non-negative functional G defined on space $\R^{p}\times \cal{P} \to$ $[0,\infty)$ is called a depth functional in regression, where $\cal{P}$ is the collection of  distribution functions on $\R^{p+1}$, if it satisfies the following four properties: 
\\[2ex]
 \noindent
 (\tb{P1}) \textbf{Invariance} (regression, scale, affine invariance)
  The functional $G$ is regression, scale and affine invariant w.r.t. a given $F_{(y,~\mathbf{x})}$ iff,  respectively,
 \bee
 G(\boldsymbol{\beta}+\mb{b};~F_{(y+\mathbf{x}'\mb{b},~ \mathbf{x})})&=&G(\boldsymbol{\beta};~F_{(y,~\mathbf{x})}), ~\forall ~\mb{b} \in \R^{p},\\[1ex]
 G(s\boldsymbol{\beta};~F_{(sy,~\mathbf{x})})&=&G(\boldsymbol{\beta};~F_{(y,~\mathbf{x})}),  ~\forall~ s (\neq 0)\in \R,\\[1ex]
 G(A^{-1}\boldsymbol{\beta};~F_{(y, ~A'\mathbf{x})})&=&G(\boldsymbol{\beta};~F_{(y,~\mathbf{x})}), ~\forall\mbox{~ nonsingular $p$ by $p$ matrix} ~ A.
 \ene
 \vskip 3mm
  \noindent
   (\tb{P2}) \textbf{Maximality at center} The functional $G$ possesses its maximum over $\boldsymbol{\beta}\in \R^{p}$ w.r.t.
  a given $F_{(y,~\mathbf{x})}$. That is, $\max_{\boldsymbol{\beta}\in \R^{p}}G(\boldsymbol{\beta};~F_{(y,~\mathbf{x})})$ exists. 
  Furthermore, it is attained at $\boldsymbol{\beta}^*$ if $\bs{\beta}^*$ 
   is the center of symmetry of $F_{(y,~\mathbf{x})})$  w.r.t. some notion of symmetry in regression.
 \vskip 3mm
  \noindent
  (\tb{P3}) \textbf{Monotonicity relative to deepest point} With respect to a maximum depth point $\bs{\beta}^*$ of the functional $G$, for any $\bs{\beta}  \in \R^p$ and $\lambda\in [0,1]$,
  \bee
  G(\lambda\boldsymbol{\beta}^*+(1-\lambda)\boldsymbol{\beta};~F_{(y,~\mathbf{x})})&\geq& G(\bs{\beta};~F_{(y,~\mathbf{x})}).
  \ene
 \vskip 3mm
 \noin
 (\tb{P4}) \textbf{Vanishing at infinity} The functional $G$ is vanishing when $\|\bs{\beta}\|\to \infty$. That is,  $\lim_{\|\bs{\beta}\|\to \infty}G(\bs{\beta};~F_{(y,~\mathbf{x})})=0$.

 \vskip 3mm

 Note that due to the reverse relationship, if the depth notion above changes to a unfitness notion, then the above four properties need obvious changes
 except the \tb{(P1)}.  Maximum in \tb{(P2)} becomes the minimum. \tb{(P3)} changes maximum to minimum and reverses the direction of the inequality. \tb{(P4)} becomes $\lim_{\|\bs{\beta}\|\to \infty}\mbox{UF}(\bs{\beta};~F_{(y,~\mathbf{x})})= \infty$. 
 \vskip 3mm
 \noindent
 The four properties above were first investigated for the simplicial depth function in Liu (1990) and formulated for general depth functions in location in ZS00. They
 have been adopted and extended for depth notions in other settings, especially for the functional data in Nieto-Reyes and Battey (2016) (NRB16) from the topological validity point of view,  for general functional data in Gijbels and Nagy (2017)(GN17), and for the
 relevance of halfspace depths in scatter, concentration and shape matrices in Paindaveine and Van Bever (2018).
 \vs
 Sophisticated discussions on the adaptations and the replacements of the four properties and the appropriateness have been given in Dyckerhoff (2004) and Serfling (2006, 2019), and in NRB16 and GN17 for functional data.
 Here for the sake of consistency and simplicity, we keep focusing on the four core axiomatic properties and make some remarks below.
 \vs

 \noin
 \textbf{Remarks 3.1}\\[1ex]
 \noindent
 (I)  \tb{(P1)} guarantees that the notion of depth in regression does not depend on the underlying coordinate system or measurement scale. This provides an advantage in the study of the depth induced functionals (estimators) by just dealing with an easily manageable special case (e.g. a spherically symmetric distribution) to cover a large class of cases (e.g. all elliptically symmetric distributions) without loss of generality (see e.g. VAR00).
 \vskip 3mm
 \noindent
 (II \tb{(P2)} says that the maximum of $G$ always exists, and it is attained at the center of symmetry w.r.t. some notion of symmetry in regression, when there is such a center.
This allows one to discuss the maximum regression depth estimating functional (or estimator in the empirical case).
  Note that the supremum of bounded $G$ always exists but not necessarily for the maximum. If \tb{(P4)} holds, one then can just focus on  bounded $\bs{\beta}$, however, since $G$ is not necessarily continuous in $\bs{\beta}$, the maximum of $G$ is not guaranteed to exist. In the empirical distribution case, however, if there are only finitely many hyperplanes
 that need to be concerned for a given depth functional, then the maximum always exists.
 \vskip 3mm
 \noindent
 (III)  \tb{(P3)} guarantees that $G(\bs{\beta};~F_{(y,\mb{x})})$ is monotonically decreasing in $\bs{\beta}$ along any ray stemming from a deepest point.  This is equivalent to the quasi-convexity of the depth functional under \tb{(P2)}, which further implies that the set of all $\boldsymbol{\beta}$ that has depth at least $\alpha ( \geq 0)$ is convex (which will be useful when studying the depth induced contours in the parameter space of $\bs{\beta}\in\R^p$), and fewer ties in depth computations of $\bs{\beta}$ (in the strictly decreasing case) will be yielded.

\vskip 3mm
 \noindent
 (IV) \tb{(P4)} dictates that when the hyperplane $H_{\bs{\beta}}$ determined by $y=\mathbf{x}'\boldsymbol{\beta}$ becomes vertical, its depth should be vanishing.
 This makes sense since when the hyperplane is vertical, it can no longer serve as an estimating functional for a linear regression parameter. It is obviously no longer useful for the prediction of future responses as well. Note that $\|\bs{\beta}\|\to\infty$ could mean (i) $|\beta_1|\to \infty$ and or (ii) $\|\bs{\beta_2}\|\to \infty$. (ii) just means the hyperplane $H_{\bs{\beta}}$ turns out to be vertical. When (i) happens, the intercept of the hyperplane $H_{\bs{\beta}}$ becomes unbounded (assume that $x_1=1$),
 the hyperplane becomes useless and its depth logically should be vanishing.
\hfill \pend

 \noindent
 \subsection{Examining depth notions}
Now that four axiomatic properties have been presented, a natural question is: do the regression depth functions
induced from the four approaches in Sections 2.3.1 to 2.3.4
satisfy all the desired properties? That is, are they really
notions of depth w.r.t.  \tb{(P1)}-\tb{(P4)}? First, let us summarize the depth functionals from these sections.\vskip 3mm

The approach in  Section 2.3.1 based on the classical objective functions induces a class of regression depth functionals, defined by
\be D_{Obj}(\bs{\beta}; F_{(y,\mb{x})},\phi, f)=\left(1+\phi\big(F_R\big)\right)^{-1},\label{D-Obj.eqn}
\ee where $R={f(r(\bs{\beta})/S(F_y))}$, $\phi$ and $f$ (objective function) are given in Section 2.2 or Example 2.1,
and $S(\cdot)$ is a scale functional that is translation invariant and scale equivalent; dividing by it achieves  scale invariance of the depth functions; it is suppressed in $D_{Obj}$.
\vs
Facility location approach  in Section 2.3.2 induces
$D_C(\bs{\beta}; F_{(y,\mb{x})})$ (or $D_C(\bs{\beta}; P)$) that  is closely related to  $\mbox{RD}_{RH}(\bs{\beta}; F_{(y,\mb{x})})$  when the distance $d$ in (\ref{D-C-regssion.eqn}) is  the $L_1$ norm.
We will only consider $D_C(\bs{\beta}; F_{(y,\mb{x})})$ with $d(x,y)=|x-y|$ as the representative for this approach.
\vs
Typical depth functionals from Section 2.3.3 (the PP approach) are  $\mbox{RD}_{RH}(\bs{\beta}; F_{(y,\mb{x})})$ and $\mbox{PRD}(\bs{\beta}; F_{(y,\mb{x})})$.
\vs
 Mizera's approach in Section 2.3.4 can recover Tukey HD in location and $\mbox{RD}_{RH}(\bs{\beta}; F_{(y,\mb{x})})$ in
regression, and its critical function (see (\ref{Mizera-cf.eqn})) could be regarded an objective function.
Its general version of tangent depth in linear regression (on page 1694) essentially recovers  $\mbox{RD}_{RH}(\bs{\beta}; F_{(y,\mb{x})})$. No distinct depth function in linear regression from this approach will be discussed here.
\vs
Consequently, in the sequel we will investigate (i) $D_{Obj}(\bs{\beta}; F_{(y,\mb{x})},\phi,f)$, (ii) $D_C(\bs{\beta}; F_{(y,\mb{x})})$, (iii) $\mbox{RD}_{RH}(\bs{\beta}; F_{(y,\mb{x})})$, and (iv) $\mbox{PRD}(\bs{\beta}; F_{(y,\mb{x})})$. 
\vs

 \vskip 3mm
\noindent
\textbf{Proposition 3.1}~ Regression depth functional  (i), (ii), (iii), and (iv)
satisfy \tb{(P1)}. \label{invariance-pro}
\vskip 3mm

\noindent
{\sc Proof}: see the Appendix.

\vskip 3mm
\noindent
\textbf{Remarks 3.2} 
\vs
(I) Without modifying the original function $A(\bs{\beta}; \mb{v})$ of MY93 (see (II) of Remarks 2.6), the induced depth functional (iv), PRD$(\bs{\beta};F_{(y,\mb{x})})$,
can never satisfy \tb{(P1)}. 
\vs
(II) As a by-product of \tb{(P1)}, maximum regression depth functionals induced from regression depth notions in Proposition 3.1 are \emph{equivariant} as declared in Corollary 3.1 below.
\vs
(III) Note that in (\ref{eqn.MY93}), $T_{P1}(F_{(sy, \mb{x})})=s^2T_{P1}(F_{(y,\mb{x})})$.
That is, by definition below, $T_{P1}$ is \emph{not scale equivariant}, contrary to the  popular belief in the literature.
\hfill \pend \vskip 3mm

\noindent
\textbf{Corollary 3.1} The maximum regression depth functionals $\bs{\beta}^*(F_{(y, \mb{x})})$ induced from (i), (ii), (iii) and (iv)
are {regression}, {scale}, and {affine} equivariant. That is, respectively,
\begin{align*}
 \bs{\beta}^*(F_{(y+\mb{x}'\mb{b}, ~\mb{x})})= \bs{\beta}^*(F_{(y, ~\mb{x})})+\mb{b},~\forall~\mb{b}\in \R^{p}; \\[1ex]
\bs{\beta}^*(F_{(sy, ~\mb{x})})=s \bs{\beta}^*(F_{(y, ~\mb{x})}),~\forall ~\text{scalar~} s(\neq 0) \in \R; \nonumber \\[1ex]
\bs{\beta}^*(F_{(y,~ A'\mb{x})})= A^{-1}\bs{\beta}^*(F_{(y, ~\mb{x})}),~\forall~ \text{nonsingular}~ A\in R^{ p\times p}. 
 \end{align*}
\vskip 3mm

\noindent
{\sc Proof}: it is trivial. \hfill \pend
\vskip 3mm

 If a maximum regression depth estimating functional $\bs{\beta}^*(F_{(y,\mb{x})})$ is equivariant, then it is symmetric w.r.t. $(y,\mb{x})$ in the sense that $\bs{\beta}^*(F_{(y,\mb{x})})=\bs{\beta}^*( F_{(-y,\mb{-x})})$.  
 By virtue of the Corollary, one can assume (w.l.o.g.) that $\bs{\beta}^*(F_{(y,\mb{x})})$ equals $\mb{0}$.  

\vskip 3mm

For the joint distribution $F_{(y,~\mb{x})}$ 
 and the univariate location estimating functional $T$ given in Example 2.3, $F_{(y,~\mb{x})}$  is said to be \emph{T-symmetric} about a $\bs{\beta}_0$ iff  for any $\mb{v}\in \mbs^{p-1}$
\be \hspace*{-0mm}\tb{(C0)}:~~~ T\big(F_{(y-\mb{x'}\bs{\beta}_0,~\mb{x}'\mb{v})} 
\big)=0, 
\label{T-symm.eqn}\ee
\vskip 3mm

\noindent
\tb{Remarks 3.3:} \vskip 3mm

(I) $T$-symmetric $F_{(y,~\mb{x})}$ includes a wide range of distributions. For example, if the univariate
functional $T$ is the mean functional, then this becomes the classical assumption in regression when $\bs{\beta}_0$ is the true parameter of the model:
the conditional expectation of the error term ${e}$ (that is assumed to be independent of $\mb{x}$)
 given $\mb{x}$  is zero, i.e. $$ \tb{(C1)}:~~~~ T(F_{(y-\mb{x'}\bs{\beta_0}, ~\mb{x}'\mb{v})}
 )= E(F_{(y-\mb{x'}\bs{\beta_0},~\mb{x}'\mb{v})} \big|_{\mb{x}=\mb{x_0}})=0,~\forall ~\mb{x_0}\in\R^{p},\hspace*{8mm} $$  \vskip 3mm

(II) When $T$ is the second most popular choice, the quantile functional, especially the median (Med) functional, the $T$-symmetric of  $F_{(y,~\mb{x})}$ about $\bs{\beta_0}$ is closely related to a weaker version (when $\mb{v}=(1,0,\cdots, 0)$) of the so-called \emph{regression symmetry} in RS04. Or  precisely, $$\hspace*{15mm}\tb{(C2)}:~~~~~ T(F_{(y-\mb{x'}\bs{\beta_0}, ~\mb{x}'\mb{v})}
)=\text{Med}(F_{y-\mb{x'}\bs{\beta_0}} \big|_{\mb{x}=\mb{x_0}})=0,~\forall ~\mb{x_0}\in\R^{p},\hspace*{30mm}$$ For a thorough discussion of this type of symmetry, refer to RS04. \hfill \pend
\vskip 3mm

 In the following, for $D_{Obj}(\bs{\beta}; F_{(y,\mb{x})},\phi,f)$, we consider only the combinations $\phi=\bs{\mu}$, (a) $f(x)=x^2$ (the case (I) of Example 2.1) and (b)  $f(x)=|x|$ (the case (II) of Example 2.1).
\vskip 3mm
\noindent
\textbf{Proposition 3.2} ~Regression depth function (i), (ii), (iii), and (iv) satisfy \tb{(P2)} in the following sense.\vskip 3mm

(a) The maximum of regression depth (i) (i.e. $D_{Obj}(\bs{\beta}; F_{(y,\mb{x})},\phi,f)$) exists
 and is attained at $\bs{\beta_0}\in \R^p$ if $\phi=\bs{\mu}$, $f(x)=x^2$ and \tb{(C1)} holds or if  $\phi=\bs{\mu}$, $f(x)=|x|$ and \tb{(C2)} holds.
\vskip 3mm
(b) The maximum of regression depth (ii) (i.e. $D_C(\bs{\beta}; F_{(y,\mb{x})})$)  exists if \tb{(A)} holds   
 and is attained at a bounded $\bs{\beta_0}\in \R^p$.\vskip 3mm
(c) The maximum of regression depth (iii)  (i.e. $\mbox{RD}_{RH}(\bs{\beta}; F_{(y,\mb{x})})$) exists if \tb{(A)} holds
 and is attained at $\bs{\beta_0}\in \R^p$ if \tb{(C2)} holds.
\vs
(d)  The maximum of regression depth (iii) (i.e. ${\mbox{PRD}}(\bs{\beta}; F_{(y,\mb{x})}, T)$) exists  
 and is attained at $\bs{\beta_0}\in \R^p$ if  \tb{(C0)} holds.
\vskip 3mm

\noindent
{\sc Proof}: see the Appendix.

\vskip 3mm
\noin
\tb{Remarks 3.4}\vskip 3mm

(I) Part (a) of the Proposition could be extended to cover more cases. If functional $\phi$ has the ``monotonicity" property ($\phi(F_{R1})\leq \phi(F_{R2})$ if $R1\leq R2$) and $f(x)$ 
has the unique minimum value, 
then  existence is guaranteed. When $\phi$ is the expectation or quantile functional, then it has  monotonicity, and if $f(x)$ is even, monotonic in $|x|$ and convex, then $f(x)$
has a unique minimum value. This covers a large class of combinations of $\phi$ and $f$.
 \vskip 3mm
(II) Existence of maximum for $D_C$ and  $\mbox{RD}_{RH}$ in the Proposition is established under \tb{(A)}. 
The latter \emph{sufficient condition} excludes the discrete distributions. In the empirical case, 
existence always holds true for both, nevertheless.
\hfill \pend
\vskip 3mm
\noindent
\textbf{Proposition 3.3} ~Regression depth function (i), (iii), and (iv) satisfy \tb{(P3)} in the following sense.\vskip 3mm
(a) The regression depth (i) (i.e. $D_{Obj}(\bs{\beta}; F_{(y,\mb{x})},\phi,f)$) monotonically decreases along any ray stemming from
a deepest point if $\phi$ has the monotonicity property (i.e. $\phi(F_{R1})\leq \phi(F_{R2})$ if $R1\leq R2$),
and $f$ is quasi-convex and has a unique minimum.
\vskip 3mm
(b)  The regression depth (iii) (i.e. $\mbox{RD}_{RH}(\bs{\beta}; P)$) monotonically decreases along any ray stemming from
a deepest point if \tb{(A)} holds.
\vs
(c)  The regression depth (iv) (i.e. ${\mbox{PRD}}(\bs{\beta}; F_{(y,\mb{x})})$) monotonically decreases along any ray stemming from a deepest point.
\vs
(d)
 The regression depth (ii) (i.e. $D_C(\bs{\beta}; P)$) \emph{violates}
 \tb{(P3)} generally. 
\vskip 3mm

\vskip 3mm

\noindent
{\sc Proof}: see the Appendix
\vskip 3mm
\noin
\tb{Remarks 3.5}.
\vs
(I) When $\phi$ is the expectation or quantile functional in (a) of the Proposition, then it has the monotonicity property, and when $f$ is $x^2$ or $|x|$ or even the check function in (III) of Example 2.1,
then it again meets all the requirements in (a) of the proposition.\vs

(II) For  $\mbox{RD}_{RH}$ to meet \tb{(P3)} (or \tb{(P2)}), we have to ask for \tb{(A)} to hold .
 \tb{(P3)} always holds for $\text{\mbox{PRD}}(\bs{\beta}; F_{(y,\mb{x})})$ with $T$ in Example 2.3. 
\hfill \pend
\vskip 3mm
\vskip 3mm
\noindent
\textbf{Proposition 3.4} Regression depth functional (i), (ii), (iii), and (iv) satisfy \tb{(P4)} in the following sense.
\vs
(a) The regression depth (i): $D_{Obj}(\bs{\beta}; F_{(y,\mb{x})},\phi,f)\to 0$ when $\|{\beta_1}\|\to \infty$ and $\|\bs{\beta_2}\|<\infty$ if $\phi(F_R)\to \infty$ as $|R|\to \infty$
and $f(x)\to \infty$ as $|x|\to \infty$.
\vskip 3mm
(b) The regression depth (ii): $D_C(\bs{\beta}; F_{(y,\mb{x})}) 
 \to 0$ when $\|\bs{\beta}\|\to \infty$ if \tb{(A)} holds.
\vskip 3mm
(c) The regression depth (iii): $\mbox{RD}_{RH}(\bs{\beta};P) \to 0$ when $\|\bs{\beta}\|\to \infty$ if \tb{(A)} holds.
\vs
(d)  The  regression depth (iv): ${\mbox{PRD}}(\bs{\beta}; F_{(y,\mb{x})}, T) \to 0$  as $\|\bs{\beta}\|\to \infty$ for $T$ in Example 2.3. 
\noindent
{\sc Proof}: see the Appendix.
\vs

\noin
\tb{Remarks 3.6}
\vs

(I) (a) is established under some assumptions on $\phi$ and $\bs{\beta}$.
 If $\phi$ is the expectation or quantile functional and $f(x)$ is even, monotonic in $|x|$ and convex, 
 then they satisfy the assumptions.
  (a) only treats one case of $\|\bs{\beta}\|\to \infty$. This is, the intercept becomes unbounded while $\|\bs{\beta_2}\|<\infty$ (as argued in (IV) of Remarks 3.1, in this case, the depth function ought to vanish). The other case of $\|\bs{\beta}\|\to \infty$ remains untouched.
\vs
(II)  (b) and (c) are established under the assumption \tb{(A)}. 
(d) holds for $\mbox{PRD}(\bs{\beta};F_{(y,~\mb{x})},T)$ with $T$ in Example 2.3 without any extra assumption. This $T$ could be the median or quantile functional or the weighted mean functional in Wu and Zuo (2009) (WZ09).
\hfill \pend

\vskip 5mm

\section{Concluding remarks}
This article extends four axiomatic properties (evaluation criteria) for location depth notions in ZS00 to depth notions in regression and discusses four general approaches for introducing notions of depth or unfitness in regression.
The latter leads to four representative depth notions: (i)
$D_{Obj}(\bs{\beta}; F_{(y,\mb{x})},\phi, f)$, (ii) $D_C(\bs{\beta}; P)$, (ii) $\mbox{RD}_{RH}(\bs{\beta}; P)$, and (iv) $\mbox{PRD}(\bs{\beta}; F_{(y,\mb{x})})$.
\vs
It characterizes (ii) and reveals that 
this depth notion in regression is not identical yet closely related to the $\mbox{RD}_{RH}$ of RH99. The latter is  contrary to a claim in the literature.

\vskip 3mm
It further investigates the
leading regression depth notions (i), (ii), (iii) and (iv) w.r.t. the evaluation criteria 
and shows that (a) $D_{Obj}(\bs{\beta}; F_{(y,\mb{x})},\phi, f)$ satisfy all the four properties under some conditions on $\phi$ and $f$,
with \tb{(P4)} proved under just one special case of $\|\bs{\beta}\|\to \infty$; (b) under \tb{(A)}, $D_C$ satisfy all (but \tb{P3}) properties;
 (c)  $\mbox{RD}_{RH}(\bs{\beta}; P)$  satisfy all the four axiomatic properties if \tb{(A)} holds; (d) ${\mbox{PRD}}(\bs{\beta}; F_{(y,\mb{x})} )$ satisfy all four properties.
 \vs
Therefore, all but Carrizosa depth (ii)  are real regression depth notions w.r.t. the four properties under those assumptions. Moreover, depth functions induced from ${\mbox{PRD}}$ are representative extensions of eminent projection depth 
 in location to regression. 
\vs
As by-product of this article, two new characterizations of RD$_{RH}$ are obtained.
One is in Proposition 2.2 and the other  in Example 2.2. The latter one turns out to be extremely helpful in studying the asymptotics of the deepest regression estimator $\bs{\beta}^*_{RD_{RH}}$ (see Zuo (2019a)).
\vs

One of the primary advantages of the notions of depth is that it can be employed directly to define
median-type deepest (or maximum depth) estimating functionals (estimators in the empirical distribution case) for
parameters in regression or location models. The most outstanding feature of the univariate median is its exceptional robustness. Do the deepest regression estimating functionals induced from  real regression depth notions here inherit this robustness property? Answers to this for  most cases of (i) and for (iii) have been given in the literature (e,g. VAR00). Encouraging answers to (iv) have been established in Zuo (2018). \vskip 3mm

Besides \tb{(P1)}-\tb{(P4)},  in evaluating and comparing the overall performance of various regression depth notions, one certainly has to further take into account the robustness and efficiency of their induced maximum depth estimators and their computability. Taking all these factors into consideration, preliminary results (see Zuo (2019b)) indicate that projection regression depth, just as its location counterpart, is competitive among leading competitors.

\section{Appendix}\label{app}
\vs
\noin
\textbf{Proof of Proposition 2.2}: \vskip 3mm 
{\sc Proof of part} (ii).  Assume that $\|\bs{\beta}\|<\infty$, we need to show that
 \be
\mbox{RD}_{RH}(\bs{\beta}; P)= \inf_{\bs{\alpha} \in S(\bs{\beta})}P\left(|r(\bs{\beta})|\leq |r(\bs{\alpha})|~\right).\label{D-C-proof.eqn}
 \ee
 Denote the angle between the hyperplane $H_{\bs{\beta}}$ (determined by $y=\mb{w'}\bs{\beta}$) and the horizontal hyperplane plane $H_h$ (determined by $y=0$) by $\theta_{\bs{\beta}}$ (consider the acute one only, hereafter). That is, $\theta_{\bs{\beta}}$ is the angle between the normal vector $(-\bs{\beta_2'}, 1)'$ and the normal vector $(\mb{0'}, 1)'$ in the $(\mb{x}', y)'$-space. Therefore, it is easy to see that $|\tan(\theta_{\bs{\beta}})|=\|\bs{\beta_2}\|$.
For any $\bs{\alpha}=(\alpha_1,\bs{\alpha}'_2)'\in S(\bs{\beta})$ ($\|\bs{\alpha}\|<\infty$) define similarly (hereafter) $H_{\bs{\alpha}}$ and $\theta_{\bs{\alpha}}$.
\vs
First we show that the LHS of (\ref{D-C-proof.eqn})) is no less than its RHS. 
Tilting ${\bs{\beta}}$ to a vertical position in Definition 2.2 means tilting $H_{\bs{\beta}}$ along a hyperline $l_v(\bs{\beta})$ which is the intersection line of $H_{\bs{\beta}}$ with some vertical hyperplane $H_v$. Let
 $P(l_v(\bs{\beta}))$ be the minimum probability mass touched by tilting $H_{\bs{\beta}}$ in the definition of $\mbox{RD}_{RH}$
 to a vatical position along $l_v(\bs{\beta})$ in two ways. Then it is readily seen that
 \be
 \mbox{RD}_{RH}(\bs{\beta};P)=\inf_{l_v({\bs{\beta}})}P(l_v(\bs{\beta})). \label{rd-proof.eqn}
 \ee

 \vs
  Let $H_{\bs{\gamma}}$ be the hyperplane  with $\theta_{\bs{\gamma}}=\arctan \left((\|\bs{\alpha_2}\|+\|\bs{\beta_2}\|)/2\right)$ which contains the hyperline $l_v(\bs{\beta})$.
 Then it is seen that $H_{\bs{\gamma}}$ is in-between  $H_{\bs{\beta}}$ and $H_{\bs{\alpha}}$ (consider again the situation that the angle formed between $H_{\bs{\beta}}$ and $H_{\bs{\alpha}}$ is acute, w.l.o.g.).
 Furthermore, points on  $H_{\bs{\gamma}}$  have the same vertical distances to  $H_{\bs{\beta}}$ and $H_{\bs{\alpha}}$. That is, $H_{\bs{\gamma}}$ bisects  the double wedge formed by $H_{\bs{\beta}}$ and $H_{\bs{\alpha}}$ (i.e. it bisects the vertical distance between the two hyperplanes).
 \vs
  Now it is not difficult to see that
 $P\left(|r(\bs{\beta})|\leq |r(\bs{\alpha})|\right)$ equals the probability mass touched by tilting
 $H_{\bs{\gamma}}$ (\emph{towards} $H_{\bs{\beta}}$ initially) along the hyperline $l_v(\bs{\beta})$ to the vertical position.
 In order to reach the infimum over $S(\bs{\beta})$,
 we need to seek $\bs{\alpha}'s$ such that the probability mass above becomes smaller.
 \vs
 Consider $\bs{\alpha}_m \in S(\bs{\beta})$  that approach $\bs{\beta}$ (or let $\theta_{\bs{\alpha}_m}\to \theta_{\bs{\beta}}$) while $H_{\bs{\alpha}_m}$ and $H_{\bs{\beta}}$ still intercept at $l_v(\bs{\beta})$ (that is, tilting $H_{\bs{\alpha}}$ {towards} $H_{\bs{\beta}}$ along $l_v(\bs{\beta})$  yields $\bs{\alpha}_m$). As $m\to \infty$, the probability mass contained in the interior of the double wedge formed between $H_{\bs{\beta}}$ and $H_{\bs{\gamma}_m}$ approaches zero and
 $P\left(|r(\bs{\beta})|\leq |r(\bs{\alpha}_m)|\right)$ decreases to the probability mass touched by tilting
 $H_{\bs{\beta}}$ to the vertical position along the hyperline $l_v(\bs{\beta})$ in one of two ways (the other way is described below).

 \vs
 Consider $\bs{\alpha}_n \in S(\bs{\beta})$ that approach $\bs{\beta}$ (or let $\theta_{\bs{\alpha}_n}\to \theta_{\bs{\beta}}$) with $H_{\bs{\alpha}_n}$ being on the \emph{other side} of $H_{\bs{\beta}}$ and still intercept at $l_v(\bs{\beta})$ (i.e. if previously $\theta_{\bs{\alpha}_m}<\theta_{\bs{\beta}}$, then $\theta_{\bs{\alpha}_n}\geq\theta_{\bs{\beta}}$ now, vice versa).
 Using the same hyperline $l_v(\bs{\beta})$ above,  one can conclude similarly that
 $P\left(|r(\bs{\beta})|\leq |r(\bs{\alpha}_n)|\right)$ decreases to the probability mass touched by tilting
 $H_{\bs{\beta}}$ to the vertical position along the hyperline $l_v(\bs{\beta})$ in the other way, as $n\to \infty$.
 \vs
 The above results imply that $\inf_{\bs{\alpha}\in S(\bs{\beta})}P\left(|r(\bs{\beta})|\leq |r(\bs{\alpha})|~\right) \leq P(l_v(\bs{\beta}))$.
  The arbitrariness of $l_v(\bs{\beta})$ (which can be any hyperline that is the intersection line of $H_{\bs{\beta}}$ and any vertical hyperplane $H_v$), in conjunction with (\ref{rd-proof.eqn}) 
 implies that $\mbox{RD}_{RH}(\bs{\beta}; P)\geq \inf_{\bs{\alpha}\in S(\bs{\beta})}P\left(|r(\bs{\beta})|\leq |r(\bs{\alpha})|~\right)$.
 \vs
 Now we show that the LHS of (\ref{D-C-proof.eqn})) is no greater than its RHS. For a given $\bs{\alpha} \in S(\bs{\beta})$, $H_{\bs{\beta}}$ intersects $H_{\bs{\alpha}}$ at a hyperline, say $l(\bs{\beta}, \bs{\alpha})$. Replace $l_v(\bs{\beta})$  with this line in the above proof,
 it is readily seen that for the given $\bs{\beta}$, $\bs{\alpha}\in S(\bs{\beta})$, and $l(\bs{\beta}, \bs{\alpha})$,
$P\left(|r(\bs{\beta})|\leq |r(\bs{\alpha})|\right)$ equals the probability mass touched by tilting
 $H_{\bs{\gamma}}$ (\emph{towards} $H_{\bs{\beta}}$ initially) along the hyperline $l(\bs{\beta}, \bs{\alpha})$ to the vertical position,
 which implies that  $P\left(|r(\bs{\beta})|\leq |r(\bs{\alpha})|\right)\geq P(l(\bs{\beta}, \bs{\alpha}))$, where $P(l(\bs{\beta}, \bs{\alpha}))$
 is again the minimum probability mass touched by tilting $H_{\bs{\beta}}$ along the hyperline $l(\bs{\beta}, \bs{\alpha})$ to the vertical position in two ways in the Definition of 2.2.   Hence, the
 $\mbox{RD}_{RH}(\bs{\beta}; P)\leq \inf_{\bs{\alpha}\in S(\bs{\beta})}P\left(|r(\bs{\beta})|\leq |r(\bs{\alpha})|~\right)$ in light of (\ref{rd-proof.eqn}).
This completes the proof of (ii)
 \vs
{\sc Proof of part} (i). Consider only the $\bs{\alpha}$ that does not belong to $S(\bs{\beta})$. Hence, $H_{\bs{\alpha}}$ is parallel to $H_{\bs{\beta}}$. Let $H_{\bs{\gamma}}$ be the hyperplane in the middle of the hyperstripe with $H_{\bs{\alpha}}$  and $H_{\bs{\beta}}$ as its two boundaries (i.e. $\theta_{\bs{\gamma}}=\arctan \left((\|\bs{\alpha_2}\|+\|\bs{\beta_2}\|)/2\right)$ ). Then
it is readily seen that
 $P\left(|r(\bs{\beta})|\leq |r(\bs{\alpha})|\right)$ equals to the probability mass carried by the closed half of the hyperstripe with $H_{\bs{\gamma}}$ and and $H_{\bs{\beta}}$ as its two boundaries.\vs
  Consider $\bs{\alpha}_n \not\in S(\bs{\beta})$ that approach $\bs{\beta}$ (or let $\alpha_{1_n}\to {\beta_1}$),
   it is readily seen that the probability mass contained in the interior of the half hyperstripe formed between $H_{\bs{\beta}}$ and $H_{\bs{\gamma}_n}$ approaches zero, and
 $P\left(|r(\bs{\beta})|\leq |r(\bs{\alpha}_n)|\right)$ decreases to 
 $P(H_{\bs{\beta}})=P(r(\bs{\beta})=0)$ as $n\to\infty$. Similarly to the proof of part (ii) above, it is readliy shown that $\inf_{\bs{\alpha} \not\in S(\bs{\beta})}P\left(|r(\bs{\beta})|\leq |r(\bs{\alpha})|\right)=P(r(\bs{\beta})=0)$. 
  On the other hand, by the proof of part (ii) above, it is
 readily seen that $\inf_{\bs{\alpha}\in S(\bs{\beta})}P\left(|r(\bs{\beta})|\leq |r(\bs{\alpha})|~\right)\geq P(r(\bs{\beta})=0).$ This completes the proof of (i).
\hfill \pend
\vskip 3mm
\vs
\noindent
\textbf{Proof of Proposition 2.3}:
\vs
(i) For any given $\bs{\beta}=(\beta_1,\bs{\beta}'_2)'$,
let the angle between the hyperplane $H_{\bs{\beta}}$ (determined by $y=\mb{w'}\bs{\beta}$) and the horizontal hyperplane plane $H_h$ (determined by $y=0$) be $\theta$. That is, $\theta$ is the angle between the normal vector $(-\bs{\beta_2'}, 1)'$ and the normal vector $(\mb{0'}, 1)'$. Therefore, it is easy to see that $|\tan(\theta)|=\|\bs{\beta_2}\|$. When $\|\bs{\beta}\|=(|\beta_1|^2+\|\bs{\beta}_2\|^2)^{1/2}\to \infty$,
assume w.o.l.g. that $|\beta_1|< \infty$ (otherwise $D_C(\bs{\beta};P)\to 0$ by definition (\ref{D-C-regssion.eqn})), then
 $|\tan({\theta})|\to \infty$, $H_{\bs{\beta}}$ turns to be vertical, which further implies by Proposition 2.2 that $D_C(\bs{\beta};P)\to 0$ since
 the closed double wedge formed by $H_{\bs{\beta}}$ and its eventual vertical hyperplane $H_v$ becomes smaller and smaller (in Lebesgue measure sense), and $H_{\bs{\beta}}$ approaches its eventual vertical hyperplane $H_v$.
  \vs
(ii) Part (i) implies that when $\|\bs{\beta}\|$ becomes unbounded, 
  the RHS of (\ref{D-C-beta^*.eqn}) cannot reach its maximum value at such $\bs{\beta}$.
 For a fixed $\bs{\alpha}$, $f(\bs{\beta}; \bs{\alpha})=P(|r(\bs{\beta})|\leq |r(\bs{\alpha})|)$ is upper
semicontinuous in $\bs{\beta}$, hence the infimum of upper semicontinuous functions $D_C(\bs{\beta};P)$ is also upper semicontinuous.
The upper semicontinuity of $D_C(\bs{\beta};P)$ in $\bs{\beta}$ over a bounded set, in conjunction with the  extreme value theorem,  yields (ii).
\hfill \pend
\vskip 3mm

\vs
\noin
\textbf{Proof of the statement in Example 2.3}\\[1ex]
Let  $\mathbf{v}=(-v_1, \mathbf{v}_2')'\in \mathbb{R}^p$, $v_1 \in \mathbb{R}$, $\mathbf{v}_2\in \mathbb{R}^{p-1}$, and $\|\mathbf{v}_2\|=1$;  $r(\boldsymbol{\beta})=y-\mathbf{w'}\boldsymbol{\beta}$ and $g(\boldsymbol{\beta},
\mb{v})=r(\boldsymbol{\beta})*((\mathbf{v_2})'\mathbf{x}-v_1)=r(\bs{\beta})\mb{w'}\mb{v}$. Here we wanted to show that
\be
\mbox{RD}_{RH}(\bs{\beta}, P)=\inf_{\|\mb{v}_2\|=1, v_1\in\R}E\left(\mathbf{I}\left(g(\boldsymbol{\beta},\mb{v})\geq 0\right) \right) \label{appendix-hd.eqn}
\ee
That is, the RHS above is equivalent to (\ref{RS04.eqn}).

\vskip 3mm
(i) 
Let us just focus on $\mb{w'}\mb{v}\geq  0$ (the case $\mb{w'}\mb{v}\leq 0$ can be treated similarly).
Since $\mb{w'}\mb{v}\geq 0$ is equivalent to $\mathbf{x'}(\mathbf{v_2})-v_1 \geq 0$,
the latter represents a closed halfpalne $H_{\mb{x}}(v_1,\mb{v_2})$ in the $\mb{x}$-hyperplane (horizontal hyperplane $y=0$ in the $(\mb{x'},y)'$ space).
From the $(\mb{x'},y)'$ space point of view, it represents a closed vertical halfspace $D$ in $(\mb{x'},y)'$ space. The intersection of this $D$ with $\mb{x}$-hyperplane (or the vertical projection of D onto $\mb{x}$-hyperplane) results in  $H_{\mb{x}}(v_1,\mb{v_2})$.
\vskip 3mm
(ii) On the other hand, given a closed vertical halfspace $D$ in  $(\mb{x'},y)'$ space, it intercepts with the
$y=0$ hyperplane (or the $\mb{x}$-hyperplane) at a closed halfplane $H_{\mb{x}}$ in $\mb{x}$-hyperplane with its boundary a hyperline $l_{\mb{x}}$ in the $\mb{x}$-hyperplane. Call the direction in the $\mb{x}$-hyperplane  that is perpendicular to the hyperline $l_{\mb{x}}$ and pointing into the halfplane $H_{\mb{x}}$  as $\mb{v_2}$. 
Denote the distance from the origin to the point on  $\mb{v_2}$ and $l_{\mb{x}}$ 
as $v_1$;  it then follows that $D$ is equivalent to
$\mb{x'}\mb{v_2}\geq v_1$ in the  $(\mb{x'},y)'$ space. That is, $\mb{w'}\mb{v}\geq  0$.
\vskip 3mm
It is readily seen from (i) and (ii) above that the RHS of (\ref{appendix-hd.eqn}) is equivalent to (\ref{RS04.eqn}). Also, it is straightforward to see that it is equivalent to (\ref{VAR00.eqn}) under the assumptions there. 
Furthermore, it can be shown that
\be
\mbox{RD}_{RH}(\bs{\beta}, P)=\inf_{\|\mb{v}_2\|=1, v_1\in\R}E\left(\mathbf{I}\left(g(\boldsymbol{\beta},\mb{v})\geq 0\right) \right)=
\inf_{\mb{v}\in \mbs^{p-1}}E\left(\mathbf{I}\left(r(\boldsymbol{\beta})*(\mathbf{v}'\mathbf{w})\geq 0\right) \right). \label{appendix-hd-1.eqn}
\ee
\vskip 3mm
\noin
Incidentally, it is seen that
\bee
&\hspace*{-23mm}\inf_{\|\mb{v}_2\|=1, v_1\in\R}E\left(\mathbf{I}\left(g(\boldsymbol{\beta},\mb{v})\geq 0\right) \right)\hspace*{99mm}&\\[1ex]
&\hspace*{-12mm}
=\inf_{\|\mb{v}_2\|=1, v_1\in\R}E\left(\mathbf{I}\left(r(\boldsymbol{\beta})*(\mathbf{v}'\mathbf{w})\geq 0\right)
\right)\hspace*{99mm}&\\[1ex]
&=\inf_{\|\mb{v}_2\|=1, v_1\in\R}\min\left\{E\left(\mathbf{I}\left(r(\boldsymbol{\beta})*(\mathbf{v}'\mathbf{w})\geq 0\right)\right)
,
~E\left(\mathbf{I}\left(r(\boldsymbol{\beta})*(-\mathbf{v})'\mathbf{w})\geq 0\right)\right)\right\} \hspace*{52mm}&\\[1ex]
&=\inf_{\|\mb{v}_2\|=1, v_1\in\R}
\min\left\{E\left(\mathbf{I}\left(r(\boldsymbol{\beta})*(\mathbf{v}'\mathbf{w})\geq 0\right)\right),
~E\left(\mathbf{I}\left(r(\boldsymbol{\beta})*(\mathbf{v}'\mathbf{w})\leq 0\right)\right)\right\} \hspace*{56mm}&\\[1ex]
&=\inf_{\|\mb{v}_2\|=1, v_1\in \R}
\min\left\{E\left(\mathbf{I}\left(g(\boldsymbol{\beta},\mb{v})\geq 0\right)\right),
~E\left(\mathbf{I}\left(g(\boldsymbol{\beta},\mb{v})\leq 0\right)\right)\right\}\hspace*{73mm}&
\ene
Note that the RHS of the last equality is the quantity used for the empirical regression depth calculation in RH99 (up to a constant factor $n$).
\hfill
 \pend
\vskip 3mm
\vs
\noin
\textbf{Proof of Proposition 3.1:}\vskip 3mm

(i) For the $D_{Obj}(\bs{\beta}; F_{(y,\mb{x})},\phi,f)$ in Section (\ref{classical-sec}), notice the facts that
\bee
(y+\mathbf{x}'\mb{b})-\mathbf{x}'(\boldsymbol{\beta}+\mb{b})&=&y-\mathbf{x}'\boldsymbol{\beta}, \\
s*y-\mathbf{x}'(s*\boldsymbol{\beta})&=&s*(y-\mathbf{x}'\boldsymbol{\beta}), s\neq 0\\
y-\mathbf{x}'A(A^{-1}\boldsymbol{\beta})&=&y-\mathbf{x}'\boldsymbol{\beta}.
\ene
These, in conjunction with the scale equivalence of S, yield the invariance of $R={f(r(\bs{\beta})/S(F_y))}$ and of the depth function.
\tb{(P1)} follows immediately for the $D_{Obj}(\bs{\beta}; F_{(y,\mb{x})},\phi,f)$ in (\ref{D-Obj.eqn}).
\vs
(ii) By (ii) of Proposition 2.3, $D_C(\bs{\beta};P)=P(y-\mb{w'}\bs{\beta}=0)$, replacing the $\mb{x}$ in (i) above verification by $\mb{w}$, it is readily seen the \tb{(P1)} follows immediately for $D_C(\bs{\beta}; P)$.\vs
(iii)
For 
${\mbox{RD}_{RH}}(\bs{\beta}; P)$, 
in the empirical case, the fact that the latter satisfies it has already been declared in Section 2.1 of RH99.
For the general population case, note that a characterization of ${\mbox{RD}_{RH}}(\bs{\beta}; P)$ is (see the proof above)
\be
{\mbox{RD}_{RH}}(\boldsymbol{\beta}; P)= \!\!\!\!
\inf_{\|\mb{v}_2\|=1, v_1\in \R}\!\!\!\!E(I(r(\bs{\beta})*(v_1,\mb{v}_2')\mb{w}\geq0))=\!\!
\inf_{\mb{v}\in\mbs^{p-1}}\!\!E(I(r(\bs{\beta})*\mb{v}'\mb{w}\geq0)), \label{eqn.rd}
\ee
Similarly to the proof in (i), \tb{(P1)} follows immediately for ${\mbox{RD}_{RH}}(\bs{\beta}; P)$. 
\vskip 3mm

(iv) For
 $\mbox{PRD}(\bs{\beta}; F_{(y,\mb{x})})$ in (\ref{eqn.PRD}). \tb{(P1)} follows straightforwardly from
 (\ref{eqn.uf}), (\ref{eqn.UF}), and (\ref{eqn.PRD}), coupled with \tb{(A1)} and \tb{(A4)}.
\hfill \pend
\vs
\noindent
\tb{Remarks 5.1}\vs
(I)
(\ref{eqn.rd}) is one of representations of the $\mbox{RD}_{RH}$. 
Many other characterizations exist. For example, one is given in RS04 displayed in (\ref{RS04.eqn}) and another one given in VAR00 is:
\begin{eqnarray}
{\mbox{RD}_{RH}}(\bs{\beta}; P)\!\!= \!\!\!\inf_{\mb{u}\in\R^{p-1},~v\in\R } \Big\{ P\left(r(\bs{\beta})> 0 \cap \mb{x'}\mb{u}<v\right)
 + P\left(r(\bs{\beta})< 0\cap \mb{x'}\mb{u}>v\right)\Big\}. &&\hspace*{3mm} \label{VAR00.eqn}
\end{eqnarray}
They assumed that $P(\mb{x'}\mb{u}=v)=0$ (and  implicitly assumed that  \tb{(A0):}  $P(r(\bs{\beta})=0)=0$).
\vskip 3mm

(II) Another representation of the $\mbox{RD}_{RH}$ given in AMY02 is displayed in (\ref{amy02-d1.eqn}),
  which is slightly more general than (\ref{VAR00.eqn}) but again also implicitly made the assumptions above. 
  The latter implies that these representations  are valid only for regression lines or hyperplanes that do not contain any probability mass. The  empirical version of (\ref{amy02-d1.eqn}) was also given on page 158 of  Maronna, Martin, and Yohai (2006) (MMY06).\vskip 3mm

(III) Empirical versions of the regression depth of RH99 and its relationship to the location (halfspace) depth were also extensively investigated in Mizera (2002) (page 1689-1690).\vskip 3mm

(IV)   Another empirical version (which actually is slightly different from $\mbox{RD}_{RH}$) was given in Bai and He (1999):
\bee \mbox{RD}_{RH}(\boldsymbol{\beta},\mathbf{Z_n})\!\!=\!\!
\inf_{\|u\|=1,~ v \in \R} \min \bigg\{\sum_{i=1}^nI(r_i(\boldsymbol{\beta})(\mathbf{u}'\mathbf{x}_i-v)>0), 
\sum_{i=1}^nI(r_i(\boldsymbol{\beta})(\mathbf{u}'\mathbf{x}_i-v)<0) \bigg\}, &&
\ene
where $y_i= \beta_0+\mathbf{x}'_i \boldsymbol{\beta_1}+e_i$, $\boldsymbol{\beta}'=(\beta_0, \boldsymbol{\beta_1}')\in \R^{p}$, $\mathbf{x_i}\in \R^{p-1}$, $r_i(\boldsymbol{\beta})=y_i-(1, \mathbf{x_i}')\boldsymbol{\beta}$, and $Z_n=\{(\mathbf{x_i}, y_i), i=1,\cdots, n\}$.
They again implicitly assumed that \tb{(A0)} hold  
 and $P(\mb{x'}\mb{u}=v)=0$.
\hfill \pend

\vs
\vs
\noin
\textbf{Proof of Proposition 3.2:}\vskip 3mm 
(a) In  light of (\ref{D-Obj.eqn}), the existence of maximizer of $D_{Obj}$ is equivalent to the existence of the minimizer of $\phi(F_R)$, where $R=f\left(r(\bs{\beta})/S(F_y)\right)$.  The latter holds true by virtue of
the property (``monotonicity") of the given functional $\phi$ (i.e. $\phi(F_{R1})\leq \phi(F_{R2})$ if $R1\leq R2$) and the unique minimizer $0$ of the given $f$ with the minimum value $0$. Under \tb{(C1)} or \tb{(C2)}, it is readily seen that the maximizer is $\bs{\beta_0}$, in the respective cases.
\vskip 3mm
(b) For $D_c(\bs{\beta}; P)$, \textbf{P2} follows directly from Proposition 2.3.\vs
(c) Following the proof of Proposition 2.3, it can show similarly that under the given condition $P(H_v)=0$, $\mbox{RD}_{RH}(\bs{\beta};P)\to 0$ when $\|\bs{\beta}\| \to \infty$ for $\bs{\beta}=(\beta_1,\bs{\beta_2}')'$ with bounded $\beta_1$. For $|{\beta_1}|\to\infty$ case,
we have to adopt the slightly modified definition for $\mbox{RD}_{RH}$, as done in (\ref{D-C-regssion.eqn}) for $D_C$. 
That is, $\mbox{RD}_{RH}(\bs{\beta};P)\to 0$ when $|{\beta_1}|\to\infty$. 
Following the arguments given in (ii) of Proposition 2.3, we see that the maximum of  $\mbox{RD}_{RH}(\bs{\beta};P)$ exists and  is attained at a bounded $\bs{\beta}^*$ (note that $\mbox{RD}_{RH}(\bs{\beta};P)$ is upper semi-continuous).
Now, if \tb{(C2)} holds, i.e. $F_{(y,\mb{x})}$ is regression symmetric about the $\bs{\beta_0}$, then by Theorem 3 of RS04
$${\mbox{RD}}_{RH}(\bs{\beta_0})=\frac{1}{2}+\frac{1}{2}P(y-\mb{w'}\bs{\beta_0}=0), $$
which is the maximum possible depth value for all $\bs{\beta}\in \R^{p}$ in this case.
\vskip 3mm

(d) We have to show that (a) the depth value can not be maximized when the norm of $\bs{\beta}\in \R^{p}$  becomes unbounded and (b) within the set of bounded $\bs{\beta}\in \R^{p}$, there exits a $\bs{\beta_0}$ which can attain the maximum depth value.
For (a), by  
Lemma 5.1 (given below), 
one immediately sees that $\mbox{PRD}(\bs{\beta})\to 0$ as $\|\bs{\beta}\|\to \infty$.
For (b), first, the continuity of $\mbox{PRD}(F_{(y, \mb{x})}; \bs{\beta})$ in
$\bs{\beta}$ follows directly from the (\tb{A3}) and the property of supremum; second, by the extreme value theorem,
the existence of a bounded maximizer $\bs{\beta_0}$ is guaranteed. 
When $F_{(y,\mb{x})}$ is $T$-symmetric about $\bs{\beta_0} \in \R^{p}$,  (iii) of Lemma 5.1 yields the desired result.
\hfill \pend
\vs
\noin
A  function $f$ from $\R^d\to \R$ is \emph{quasi-concave} if
$f(\lambda x + (1-\lambda)y) \geq  \min \{ f(x), f(y)\},$~  $\forall ~\lambda \in [0,1] ~\text{and} ~x, ~  y \text{~in}~ \R^d ~ (d\geq 1) $.
For the distribution $F_X$ of  any random vector $X$,  denote its empirical version by $F^n_X$.
 \vs
 \noin
\tb{Lemma 5.1}  The projection regression depth PRD$(\bs{\beta};F_{(y,\mb{x})})$ in (\ref{eqn.PRD}) is
 \begin{itemize}
  \vspace*{-1mm}
 \item[] (i) affine invariant, 
 quasi-concave and continuous in $\bs{\beta}$,
 \vspace*{-1mm}
 \item[] (ii) vanishing when $\|\bs{\beta}\|\to \infty$,
 \vspace*{-1mm}
 \item[] (iii)  maximized at the center $\bs{\beta_0}$ of T-symmetric $F_{(y,\mb{x})}$,
 \vspace*{-1mm}
 \item[] (iv) continuous in $F_{(y,\mb{x})}$ in the sense that PRD$(\bs{\beta};F^n_{(y,\mb{x})}) \xrightarrow{m} \text{\mbox{PRD}}(\bs{\beta};F_{(y,\mb{x})})$ in the same mode as $F^n_{(y,\mb{x})} \xrightarrow{m} F_{(y,\mb{x})}$ when $n\to \infty$, provided that (for a given $\bs{\beta}$)
      \\[.5ex] (a)
      $T\big(F^n_{(y-\mb{x'}\bs{\beta})/\mb{x}'\mb{v}}\big)\xrightarrow{m} T(F_{(y-\mb{x'}\bs{\beta})/\mb{x}'\mb{v}})$ uniformly in $\mb{v}\in\mbs^{p-1}$ and
      $S(F^n_y) \xrightarrow{m} S(F_y)$,
      \\[.5ex] (b) $\sup_{\|\mb{v}\|=1}|T(F_{(y-\mb{x'}\bs{\beta})/\mb{x}'\mb{v}})|<M_T$, and $\inf_n S(F^n_y)>M_S>0$,\\[.5ex]
    where convergence mode ``m" could be in $o_P(1)$, $o(1)$ a.s., or in $O_P(n^{-1/2})$.
 \end{itemize}
 \vs
 \noin
 \tb{Proof of Lemma 5.1}\vs
 (i) This is a straightforward verification, by \tb{(A1)}, \tb{(A3)}, and (\ref{eqn.uf}), (\ref{eqn.UF}), and (\ref{eqn.PRD}).
 \vs
 (ii) In  light of \tb{(A1)}, it is readily seen that   $T(F_{(y-\mb{x'}\bs{\beta})/\mb{x}'\mb{v}})=T(F_{(y/\mb{x'}\mb{v}})-\|\bs{\beta}\|$ for $\bs{\beta}\neq 0$ and $\mb{v}=\bs{\beta}/\|\bs{\beta}\|$.
 This, in conjunction with \tb{(A2)}, (\ref{eqn.uf}), (\ref{eqn.UF}), and (\ref{eqn.PRD}), yields  PRD$(\bs{\beta}; F_{(y,\mb{x})})\to 0$ as $\|\bs{\beta}\|\to \infty$.
 \vs
 (iii) In virtue of the definition of T-symmetric about $\bs{\beta_0}$ in (\ref{T-symm.eqn}), (\ref{eqn.uf}), (\ref{eqn.UF}), and (\ref{eqn.PRD}), one sees that
 PRD$(\bs{\beta_0}; F_{(y,\mb{x})})$ attains its maximum possible value $1$.
 \vs
 (iv) Write $G$ for PRD, and in light of (\ref{eqn.PRD}), (\ref{eqn.UF}), and (\ref{eqn.uf}), a simple derivation leads to
\bee
|G(\bs{\beta}; F^n_{(y,\mb{x})})-G(\bs{\beta}; F_{(y,\mb{x})})| 
\leq \sup_{\|\mb{v}\|=1}\big|\text{UF}_{\mb{v}}(\bs{\beta};F^n_{(y,\mb{x})}, T)-\text{UF}_{\mb{v}}(\bs{\beta};F_{(y,\mb{x})}, T)\big|\hspace*{17mm}&\\[2ex]
\leq
\sup_{\|\mb{v}\|=1}\frac{D(T_n)S(F_y)+D(S_n)|T(F_{(y-\mb{x'}\bs{\beta})/\mb{x}'\mb{v}})|}{S(F^n_y)S(F_y)}\hspace{16mm}&\\[2ex]
\leq \frac{1}{M_S}\sup_{\|\mb{v}\|=1}D(T_n)+\frac{M_T}{M_S S(F_y)}D(S_n)\hspace*{34mm}&\\[-1ex]
\ene
by the given (b), where $D(T_n)=|T(F^n_{(y-\mb{x'}\bs{\beta})/\mb{x}'\mb{v}})- T(F_{(y-\mb{x'}\bs{\beta})/\mb{x}'\mb{v}})|$ and $D(S_n)=|S(F_y^n)-S(F_y)|$. This, in conjunction with the given (a), leads immediately to (iv).
 \hfill \pend
 \vs
 \noin
 \tb{Remarks 5.2}\vs
 (I) The assumption (a) in (iv) of the Lemma holds for classical location estimating functionals $T$  such as Med functional or trimmed and winsorized mean functionals of WZ09 and for  $S$ such as MAD or trimmed and winsorized standard deviations functionals (Wu and Zuo (2008)). Uniformity in $\mb{v} \in\mbs^{p-1}$ can usually be established via the II.4. and II.5. of Pollard (1984) .
 \vskip 3mm

(II)  The assumption (b) in (iv) of the Lemma holds true for $T$ and $S$ above, as long as $S(F_y)>\delta>0$ and $|T(F_{(y-\mb{x'}\bs{\beta})/\mb{x'}\mb{v}})|<M_{\mb{v}}<\infty$ for any $\mb{v}\in\mbs^{p-1}$.  
 \hfill \pend
\vs
\vs
\noindent
\textbf{Proof of Proposition 3.3}:\vs

It is readily seen that to prove \tb{(P3)}, it suffices to show that (1) there exists a deepest point $\bs{\beta}_0$ of $G(\bs{\beta};P)$, and (2) the regression depth function $G(\bs{\beta};P)$ is quasi-concave in $\bs{\beta}$.\vs

(a) For regression depth (i)(i.e. $D_{Obj}(\bs{\beta}; F_{(y,\mb{x})},\phi,f)$), (1) follows from the given condition and (I) of Remarks 3.4. For (2), in light of (\ref{D-Obj.eqn}), we only need to show that $\phi(F_R)$ is quasi-convex in $\bs{\beta}$ with $R=f(r(y-\mb{x}'\bs{\beta})/S(F_y))$. The latter follows from the quasi-convexity of $f$ and the ``monotonicity" of $\phi$.\vs

(b) For regression depth (iii) (i.e. $\mbox{RD}_{RH}$), (1) follows from (ii) of Proposition 3.2 and (2) follows directly from (iii) of Lemma 5.2 given below.

\vskip 3mm
(c) For the projection regression depth functional, (1) follows from the  (i) and (ii) of Lemma 5.1. For (2), in virtue of definition (\ref{eqn.PRD}),  it is  seen that it suffices to show that $\mbox{UF}(F_{(y,~\mathbf{x})};\boldsymbol{\beta}, T)$ is quasi-\emph{convex} in $\boldsymbol{\beta}$. By (\tb{A3}), coupled with (\ref{eqn.uf}) and (\ref{eqn.UF}), it is readily  seen that for any $\bs{\beta}=\lambda \bs{\beta_1}+(1-\lambda)\bs{\beta_2}, \lambda \in [0,1]$
\bee
\mbox{UF}(\boldsymbol{\beta};~F_{(y,~\mathbf{x})}, T) &\leq& \max\left\{\mbox{UF}(\boldsymbol{\beta}_1; ~F_{(y,~\mathbf{x})}, T);~~ \mbox{UF}(\boldsymbol{\beta}_2;~F_{(y,~\mathbf{x})}, T) \right\},
\ene
This completes the proof of (\tb{P3}) for $\mbox{PRD}(\boldsymbol{\beta}; ~F_{(y,~\mathbf{x})}, T)$.
\vskip 3mm

(d) Let $\lambda \in (0, 1)$ is fixed. For given $\bs{\beta}_1$ and $\bs{\beta}_2(\not= \bs{\beta}_1)$ in $\R^p$, assume that
$P(r(\bs{\beta}_i) = 0) = 1/2$ , $i = 1, 2$. Then we have that $D_C(\lambda\bs{\beta}_1 + (1-\lambda)\bs{\beta}_2; P) = 0$  and
$D_C(\bs{\beta}_i; P) = 1/2$ by Proposition 2.2. This implies that
\[ D_C(\lambda\bs{\beta}_1 + (1-\lambda)\bs{\beta}_2; P) < \min
\{D_C(\bs{\beta}_1; P); D_C(\bs{\beta}_2; P)\}=D_C(\bs{\beta}_2;P),
\]
That $D_C(\bs{\beta};P)$ attains it maximum value at $\bs{\beta}_1$ yields the desired result.
\hfill \pend
\vs
\noin
\tb{Lemma 5.2}.  $\mbox{RD}_{RH}(\bs{\beta}; P)$ of RH99 is
\vspace*{-0mm}
\begin{itemize}
\item[](i) upper semicontinuous in $\bs{\beta}$, and continuous in $\bs{\beta}$ if the density of $P$ exists
and discontinuous in $\bs{\beta}$ generally;
\item[] (ii)  continuous in $P$ in the sense that $\mbox{RD}_{RH}(\bs{\beta}; Q_n)$ converges to $\mbox{RD}_{RH}(\bs{\beta}; P)$ in the same mode (in distribution, in probability, with probability one) as $Q_n$ converges to $P$; If $Q_n$ is the empirical version $P_n$ of $P$,
    then $\mbox{RD}_{RH}(\bs{\beta}; P_n)$ converges to $\mbox{RD}_{RH}(\bs{\beta}; P)$ almost surely and uniformly in $\bs{\beta}\in\R^p$.
\item[] (iii)  quasi-concave in $\bs{\beta}\in \R^{p}$.
\end{itemize}
\vskip 3mm
\noindent
\tb{Proof of Lemma 5.2}:\vskip 3mm

(i) For fixed  $\mb{v}$, $f(\bs{\beta}; \mb{v})=P(r(\bs{\beta})*\mb{v}'\mb{w})$ in (\ref{eqn.rd}) is upper
semicontinuous in $\bs{\beta}$, hence the infimum of upper semicontinuous functions $\mbox{RD}_{RH}(\bs{\beta};P)$ is also upper semicontinuous. If a density of $P$ exists, then $f(\bs{\beta}; \mb{v})$ is continuous in  $\bs{\beta}$ and
so is the infimum of continuous functions $\mbox{RD}_{RH}(\bs{\beta};P)$.
 We focus on the discontinuity part. Suppose that the distribution of $(y,\mb{x})$ has its entire probability mass on the hyperplane determined by $y=\mb{w}'\bs{\beta}_0$ for some $\bs{\beta}_0 \in \R^{p}$, and any hyperline contains zero probability mass; then $\mbox{RD}_{RH}(\bs{\beta}_0; P)=1$, and $\mbox{RD}_{RH}(\bs{\beta}; P)=0$ for any $\bs{\beta}(\neq \bs{\beta}_0)  \in \R^{p}$.
Thus, when $\bs{\beta}$ approaches $\bs{\beta}_0$,  $\mbox{RD}_{RH}(\bs{\beta}; P)$ can never approach $\mbox{RD}_{RH}(\bs{\beta}_0; P)$.
\vskip 3mm

(ii) First part follows directly from the characterization of $\mbox{RD}_{RH}(\bs{\beta}; P)$ given in (\ref{eqn.rd}) and the continuity of the infimum function; the second part follows from standard empirical process theory, such as Pollard (1984). \vskip 3mm

(iii) Let $\boldsymbol{\beta}_1, \boldsymbol{\beta}_2 \in \R^p$ and $\lambda \in [0,1]$,
and $\boldsymbol{\beta}:=\lambda \boldsymbol{\beta}_1+(1-\lambda)\boldsymbol{\beta}_2$.
Let $H_{\bs{\beta}}$ be the hyperplane determined by $y=\mb{w}'\bs{\beta}$,  and $a=\min\{\mb{w}'\bs{\beta_1}, \mb{w}'\bs{\beta_2}\}$, $b=\max\{\mb{w}'\bs{\beta_1}, \mb{w}'\bs{\beta_2}\}$.  Denote by $W(H_{\bs{\beta_1}}, H_{\bs{\beta_2}})=\{(\bs{x}',y): \bs{x}\in \R^{p-1}, y\in[a, b] \}$ the closed double wedge formed by  two
hyperplanes $H_{\bs{\beta_1}}$ and $H_{\bs{\beta_2}}$ (assume w.l.o.g. that $H_{\bs{\beta_1}}$ is not parallel to $H_{\bs{\beta_2}}$). \vskip 3mm

By 
Definition 2.2, $\mbox{RD}_{RH}(\bs{\beta};P)$ is the minimum probability mass that needs to pass when $H_{\bs{\beta}}$ is tilted into a vertical position.  Notice that the position of $H_{\bs{\beta}}$ is in-between that of $H_{\bs{\beta_1}}$ and $H_{\bs{\beta_2}}$, it is readily seen that
\bee
\mbox{RD}_{RH}(\lambda \boldsymbol{\beta}_1+(1-\lambda)\boldsymbol{\beta}_2; P)& \geq & \min\big\{ \mbox{RD}_{RH}(\bs{\beta_1}; P), \mbox{RD}_{RH}(\bs{\beta_2};P)\big \}\\[1ex]
&& +\min\big\{P(W(H_{\bs{\beta}}, H_{\bs{\beta_1}})),P(W(H_{\bs{\beta}}, H_{\bs{\beta_2}}))\big\}\\[1ex]
&\geq&  \min\big\{ \mbox{RD}_{RH}(\bs{\beta_1}; P), \mbox{RD}_{RH}(\bs{\beta_2};P) \big\}
\ene

This completes the proof of part (iii). 
 \hfill \pend\vskip 3mm

\vs
\noin
\textbf{Proof of Proposition 3.4:}\vskip 3mm
(a) By  virtue of its definition (\ref{D-Obj.eqn}), it suffices to show that $\phi(F_R)\to \infty$ when $\|\bs{\beta}\|\to \infty$ with $R=f\big(r(\bs{\beta})/S(F_y)\big)$. The latter, in light of given conditions,  follows if we can show that
$|r(\bs{\beta})|\to \infty$ when $|{\beta_1}|\to \infty$. Note that $S(F_y)$ is a fixed positive number and $|r(\bs{\beta})|\geq|\beta_1|-|y-\bs{\beta_2}'\mb{x}|\geq |\beta_1|-|\bs{\beta_2}'\mb{x}|-|y|\geq
|\beta_1|-|y|-\|\bs{\beta_2}\|\|\mb{x}\|\to \infty$ with probability one. This implies that $|R|\to \infty$ by virtue of the given condition on $f$, which in turn implies that $\phi(F_R)\to \infty$.

\vskip 3mm
(b)  $D_C(\bs{\beta},P)$  satisfies \tb{P4} follows directly from the Proposition 2.3. 
\vskip 3mm
(c) RD$_{RH}(\bs{\beta}; P)$  satisfies \tb{P4} has been proved in (c) of the proof of Proposition 3.2.
\vs
(d) This part was given in (ii) of Lemma 5.1.
\hfill \pend
\section*{Acknowledgements}
The author thanks Hanshi Zuo and Professors Emeritus Hira Koul and James Stapleton for their careful English proofreading of the manuscript and insightful comments and suggestions which have led to distinct improvements.
{\small


\begin{thebibliography}{9}
\bibitem{AMY02} Adrover, J., Maronna, R. and Yohai, V. (2002), ``Relationships between maximum depth and
projection regression estimates", \emph{J. Statist. Plann. Inference}, 105, 363-375

\bibitem{AR11} Agostinelli, C., and Romanazzi, M. (2011), ``Local depth", \emph{J. Statist. Plann. Inference}, 141, 817-830.

\bibitem{BH99}Bai,Z.\ D.\, and He, X. (1999), ``Asymptotic distributions of the maximal depth regression and multivariate location'',  \emph{ Ann. Statist.}, Vol. 27, No. 5, 1616--1637.
\bibitem{BT74} Beaton, A. E., and Tukey, J. W. (1974), ``The fitting of power series, meaning
polynomials, illustrated on band-spectroscopic data", \emph{Technometrics}, 16, 147--185.
\bibitem{B95} Billingsley, P. (1995), ``Probability and Measure", John Wiley \& Sons, New York.
\bibitem{CN88}Caplin, A.\  and  Nalebuff, B.\ (1988), ``On 64\%-majority rule'', \emph{Econometrica},56, 787--814.

\bibitem{CN91a} Caplin, A.\ and  Nalebuff, B.\ (1991a), ``Aggregation and social choice: A mean voter theorem'', \emph{Econometrica}, 59, 1--23.

\bibitem{CN91b}Caplin, A.\ and  Nalebuff, B.\ (1991b), ``Aggregation and imperfect competition: On the existence of equilibrium'', \emph{Econometrica}, 59, 25 – 59.
\bibitem{C96} Carrizosa, E. (1996), ``A characterization of halfspace depth", \emph{J. Multivariate Anal.}, 58 21-26.
\bibitem{CC14} Chakraborty, A.\ and Chaudhuri, P. (2014), ``The spatial distribution in infinite dimensional spaces and related quantiles and depths",
{\it Ann. Statist.}, 42(3), 1203-1231.
\bibitem{CO11} Chebana, F.\ ,  Ouarda, T.\ B.\ M.\ J.\ (2011), ``Depth-based multivariate descriptive statistics with hydrological applications",  \emph{Journal of Geophysical Research}, 116, D10120, 10.1029/2010JD015338.

\bibitem{CO08} Chebana, F.\ ,  Ouarda, T.\ B.\ M.\ J.\ (2008), ``Depth and homogeneity in regional flood frequency analysis'', \emph{Water Resources Research}, 44, W11422, doi:10.1029/2007WR006771.

\bibitem{CGR17}Chen, M., Gao, C. and Ren, Z. (2018), ``Robust covariance and scatter matrix estimation
under Huber's contamination model", \emph{Ann. Statist.},
46(5), 1932-1960.
 \bibitem {C 17} Chernozhukov, V.\ , Galichon, A.\ , Hallin, M.,  and Henry, M. (2017), ``Monge--Kantorovich depth, quantiles, ranks and signs",\emph{ Ann. Statist.}, 45(1), 223-256.

\bibitem{CHSV14}Claeskens, G., Hubert, M., Slaets, L. and Vakili, K. (2014), ``Multivariate functional
halfspace depth". \emph{J. Amer. Statist. Assoc.}, 109, 411-423.

\bibitem{CL08} Cui, X.\ , Lin, L.\ , and   Yang, G.\ (2008), ``An Extended Projection Data Depth and Its Applications to Discrimination", \emph{Communications in Statistics - Theory and Methods}, 37(14), 2276--2290, DOI: 10.1080/03610920701858396.

 \bibitem{DS10} Dang, X., and Serfling, R.\ (2010), ``Nonparametric depth-based multivariate outlier identifiers, and masking robustness properties", \emph{J. Statist. Plann. Inference}, 140, 198-213.
\bibitem{D82}  Donoho, D.\ L.\ (1982), ``Breakdown properties of multivariate location estimators", PhD Qualifying
 paper, Harvard Univ.
 \bibitem{DG92} Dohono, D. L. and Gasko, M. (1992), ``Breakdown properties of location estimates based on
halfspace depth and projected outlyingness", \emph{Ann. Statist.}, 20 1803-1827.
\bibitem{D16} Dyckerhoff, R. (2004), ``Data depths satisfying the projection
property", Allgemeines Statistisches Archiv 88, 163-190.
\bibitem{E03} Eisenhauer, J.\ G. (2003), ``Regression through the Origin", \emph{Teaching Statistics}, 25(3), 76-80.
\bibitem{FGGM08}Febrero,  M.\ ,  Galeano, P.\ , and Gonzalez-Manteiga, W.\  (2007), ``Outlier detection in functional data by depth measures, with application to identify abnormal NOx levels", \emph{Environmetrics},  DOI: 10.1002/env.878.
\bibitem{GP05}  Ghosh A.\ K.\, and  Chaudhuri, P.\ (2005), ``On data depth and distribution free discriminant
analysis using separating surfaces". \emph{Bernoulli} 11, 1–27.
\bibitem{GN17} Gijbels, I. and Nagy, S. (2017), ``On a general definition of depth
for functional data", \emph{ Statist. Sci.}, 32(4), 630-639.

\bibitem{HPS10} Hallin, M., Paindaveine, D. and \"{S}iman. M. (2010), ``Multivariate quantiles and multipleoutput regression quantiles: From L1 optimization to halfspace depth" (with discussion),  \emph{Ann. Statist.}, 38 (2), 635-703.
\bibitem{Ham74} Hampel, F. R. (1974), ``The influence curve and its role in robust estimation", {J.\
Am.\ Stat.\ Asoc.}, 69, 383--393.

\bibitem{H00} Hoberg, R.\ (2000), ``Cluster analysis based on data depth'', \emph{In Data Analysis, Classification
and Related Methods (H. Kiers, J. P. Rasson, P. Groenen and M. Schader, eds.)},
17--22. Springer, Berlin.

\bibitem{H64} Huber, P.\ J.\ (1964), ``Robust estimation of a location parameter'', \emph{Ann. Math. Statist.}, 35 73-101
\bibitem{H73} Huber, P.J. (1973), ``Robust regression: Asymptotics, conjectures and Monte Carlo",  \emph{Ann. Statist.},, 1, 799-821.
\bibitem{HRS15}Hubert, M.\ , Rousseeuw, P.\ J.\ , and Segaert, P. (2015), ``Multivariate functional outlier detection," \emph{Statistical Methods \& Applications},  vol. 24(2), pages 177-202.
\bibitem{HRV99} Hubert, M., Rousseeuw, P. J. and Van Aelst, S. (1999), ``Similarities between location
depth and regression depth". In \emph{Statistics in Genetics and in the Environmental Sciences}
(L. Fernholz, S. Morgenthaler and W. Stahel, eds.) 159-172. Birkhäuser, Basel.
\bibitem{Ja72} Jaeckel, L. A. (1972), ``Estimating regression coefficients by minimizing the
dispersion of residuals", \emph{Ann. Math. Stat.}, 5 , 1449--1458.
\bibitem{J04} J$\ddot{o}$rnsten, R. (2004),  ``Clustering and classification based on the L1 data depth", \emph{J. Multivariate Anal.}, 90(1), 67--89.

\bibitem{Ju71} Jure\v{c}kov\'{a}, J. (1971), ``Nonparametric estimate of regression coefficients", \emph{Ann. Math. Stat.}, 42, 1328--1338.
\bibitem{KB78} Koenker, R., and Bassett, G.\ J.\ (1978), ``Regression Quantiles", \emph{Econometrica}, 46, 33--50.
\bibitem{KM97} Koshevoy, G.\ , and Mosler, K. (1997), ``Zonoid trimming for multivariate distributions",
\emph{Ann. Statist.}, 25 1998–2017.
\bibitem{K70} Koul, H. (1970), `` Some convergence theorems for ranks and weighted empirical cumulatives". \emph{Ann. Math. Statist.}, 41, 1768-1773.
\bibitem{K70} Koul, H. (1971), `` Asymptotic behavior of a class of confidence regions based on ranks in regression", \emph{Ann. Math. Statist.}, 42, 466-476.
\bibitem{LM14}Lange, T., Mosler, K.\,  Mozharovskyi, P. (2014), ``Fast nonparametric classification based on data depth", \emph{Statistical Papers}, 55(1), 49--69.
\bibitem{LL04} Li, \ J. and Liu, R.\ Y.\ (2004), ``New nonparametric tests of multivariate locations and scales using data depth", \emph{Statist. Sci.}, 686-696.
\bibitem{LCL12} Li, \ J., Cuesta-Albertos, J.\ A.\ , and  Liu, R.\ Y.\ (2012), ``DD-Classifier: Nonparametric Classification Procedure Based on DD-Plot", \emph{J. Amer. Statist. Assoc}, 107, 737–753.
\bibitem{L90} Liu, R.\ Y.\ (1990), ``On a notion of data depth based on random simplices'',
\emph{Ann. Statist.}, 18, 405-414, 1990.
\bibitem{L92}  Liu, R.\ Y.\ (1992), ``Data depth and multivariate rank tests'', \textit{In Y. Dodge (ed.),
L1-Statistical Analysis and Related Methods}, North-Holland, Amsterdam,
279-294.
\bibitem{L95}Liu, R.\ Y.\  (1995), ``Control charts for multivariate processes'', \emph{J. Amer. Statist. Assoc.}, 90, 1380--1387.
\bibitem{LPS99}  Liu, R.\ Y.\ Parelius, J.\ M.\ , and Singh, K.\ (1999), ``Multivariate analysis
by data depth: Descriptive statistics, graphics and inference'', \emph{Ann. Statist.}, 27, 783-858. With discussion.
\bibitem{LS93} Liu, R.\ Y.\ , and  Singh, K.\ (1993), ``A quality index based on data depth and multivariate
rank tests", \emph{J. Amer. Statist. Assoc.}, 88:252–260.
\bibitem{LZ14} Liu, X. and Zuo, Y. (2014), ``Computing halfspace depth and regression depth",
\emph{Communications in Statistics - Simulation and Computation}, 43(5), 969-985.
\bibitem{LPR09} Lopez-Pintado, S.\ and Romo, J. (2009), ``On the Concept of Depth for Functional Data," \emph{J. Amer. Statist. Assoc.},  104(486), pages 718-734.
\bibitem{17}Majumdar, S., and Chatterjee, S. (2017), ``Nonconvex penalized multitask regression using data depth-based penalties", arXiv:1610.07540v3
\bibitem{MMY06} Maronna, R.\ A., Martin, R.\ D., and Yohai, V.\ J.(2006), \emph{Robust Statistics: Theory and Methods},  John Wiley \& Sons
\bibitem{MY93} Maronna, R.\ A.,\ and Yohai, V.\ J.\ (1993), ``Bias-Robust Estimates of Regression Based on Projections'', \emph{Ann. Statist.}, 21(2), 965-990.
\bibitem{M02} Mizera, I. (2002), ``On depth and deep points: a calculus", \emph{Ann. Statist.}, 30(6), 1681--1736.
\bibitem{MM04} Mizera, I. and M\"{u}ller, C.\ H. (2004), ``Location-Scale Depth " (with discussions),  \emph{J. Amer. Statist. Assoc.}, 99(468), 949-989.
\bibitem{M02}Mosler, K.\ (2002), \textit{Multivariate Dispersion, Central Regions and Depth:
The Lift Zonoid Approach}. Springer, New York.

\bibitem{MB14} Mosler, K.\ and Bazovkin, P.\ (2014), ``Stochastic linear programming with a distortion
risk constraint", \emph{OR Spectrum}, 36 949--969.

\bibitem{MH06} Mosler, K.\ and  Hoberg, R.\ (2006), ``Data analysis and classification with the zonoid
depth",  \emph{In Data Depth: Robust Multivariate Analysis, Computational Geometry and Ap-
plications (R. Liu, R. Serfling and D. Souvaine, eds.)}, 49–59, American Mathematical Society, Providence RI.
\bibitem{NRB16} Nieto-Reyes, A. and Battey, H. (2016), ``A topologically valid definition of depth for functional data", \emph{Statist. Sci.}, 31 61-79.
\bibitem{PVB13} Paindaveine, D. and Van Bever, G. (2013), ``From Depth to Local Depth:
A Focus on Centrality",  \emph{J. Amer. Statist. Assoc.}, 108(503), 1105-1119.
\bibitem{PVB15} Paindaveine, D.\ and Van Bever, G.\ (2015), ``Nonparametrically consistent depth-based
classifiers", \emph{Bernoulli} 21 62-82.
\bibitem{PVB17} Paindaveine, D.\ and Van Bever, G.\ (2018), ``Halfspace depths for scatter, concentration and shape matrices", \emph{Ann. Statist.},
46(6B) , 3276-3307.
\bibitem{P84} Pollard, D. (1984), \emph{Convergence of Stochastic Processes}, Springer, Berlin.

\bibitem{P03}Portnoy, S. (2003), ``Censored Quantile Regression", \emph{J. Amer. Statist. Assoc.},
98, 1001--1012.
\bibitem{P12}Portnoy, S. (2012), ``Nearly root-n approximation for regression quantile processes",\emph{Ann. Statist.},  40(3), 1714--1736.
\bibitem{R84} Rousseeuw, P.\ J. (1984), "Least Median of Squares Regression," \emph{J. Amer. Statist. Assoc.}, 79, 871--880.
\bibitem{RH99} Rousseeuw, P.\ J., and Hubert, M. (1999), ``Regression depth'' (with discussion), \emph{J. Amer. Statist. Assoc.}, 94: 388--433.
\bibitem{RL87}Rousseeuw, P.\ J., and Leroy, A. (1987), \emph{Robust regression and outlier detection}, John Wiley \& Sons, Inc.
\bibitem{RS04}Rousseeuw, P.\ J., and  Struyf, A. (2004), ``Characterizing angular symmetry and
regression symmetry", \emph{J. Statist. Plann. Inference}, 122, 161-173.

\bibitem{RC80}Ruppert, D., and Carroll, R.\ J. (1980), ``Trimmed least squares estimation in the
linear model", {\it J. Am. Stat. Assoc.}, 5 , 828--838.
\bibitem{S06} Serfling, R. (2006), ``Depth functions in nonparametric multivariate
inference", \emph{In Data Depth: Robust Multivariate Analysis,
Computational Geometry and Applications}, DIMACS Ser. Discrete
Math. Theoret. Comput. Sci. 72 1–16. Amer. Math. Soc., Providence, RI.
\bibitem{S19} Serfling, R. (2019), ``Perspectives on Depth Functions on General
Data Spaces, with Consideration of the Tukey,
Projection, Spatial, `Density', `Local', and
`Contour' Depths", preprint.
\bibitem{SW14} Serfling, R., and Wang, S. (2014), ``General foundations for studying masking and swamping robustness of outlier identifiers", \emph{Statistical Methodology}, 20, 79-90
\bibitem{S81} Stahel, W.\ A.\ (1981), Robuste Schatzungen: Infinitesimale Optimalitiit und Schiitzungen von
 Kovarianzmatrizen. Ph.D. dissertation, ETH, Zurich.
 \bibitem{T75} Tukey, J.\ W. (1975),  ``Mathematics and the picturing of data'', \textit{In: James, R.D. (ed.), Proceeding of the International Congress of Mathematicians}, Vancouver 1974 (Volume 2), Canadian Mathematical Congress, Montreal, 523-531.
 \bibitem{VR00} Van Aelst, S., and Rousseeuw, P.\ J. (2000), ``Robustness of Deepest Regression", \emph{J. Multivariate Anal.}, 73, 82--106.

 \bibitem{VZ00} Vardi, Y. and Zhang, C.-H. (2000), ``The multivariate L1-median and associated data
depth", \emph{Proc. Natl. Acad. Sci.} USA, 97 1423-1426.
 \bibitem{VA11}  Velasco-Forero, S,\ and Angulo, J.\  (2011), ``Mathematical Morphology for Vector Images Using Statistical Depth", {\it In Pierre Soille, Martino Pesaresi, Georgios K. Ouzounis (Eds.)}, Mathematical Morphology and Its Applications to Image and Signal Processing, Springer, 355--366.

\bibitem{VA12} Velasco-Forero, S,\ and Angulo, J.\  (2012),  ``Random projection depth for multivariate mathematical
morphology'', \emph{IEEE Journal of Selected Topics in Signal Processing}, IEEE, 6 (7), 753-763.

 \bibitem{WS15} Wang, S. and Serfling, R. (2015), ``On masking and swamping robustness of leading nonparametric outlier identifiers for univariate data", \emph{J. Statist. Plann. Inference}, 162, 62-74.
\bibitem{WS15} Wang, S. and Serfling, R. (2018), ``On masking and swamping robustness of leading nonparametric outlier identifiers for multivariate data", \emph{J. Multivariate Anal.}, 166, 32-49.
 \bibitem{WZ08} Wu, M., and Zuo, Y. (2008), ``Trimmed and Winsorized Standard Deviations based on a scaled deviation", \emph{Journal of Nonparametric Statistics}, 20(4):319-335.
\bibitem{WZ08} Wu, M., and Zuo, Y. (2009), ``Trimmed and Winsorized means based on a scaled deviation",
\emph{J. Statist. Plann. Inference}, 139(2) 350-365.
\bibitem{YS97}  Yeh, A.\ B.\ , and Singh, K.\ (1997), ``Balanced confidence regions based on Tukey's depth and the bootstrap", \emph{J Roy. Statist. Soc. Ser. B}, 59 639-652.
\bibitem{Z98} Zuo, Y. (1998), ``Contributions to the Theory and Applications of Statistical
Depth Functions,"  Ph.D. thesis, The University of Texas at Dallas, Ann
Arbor, MI: ProQuest LLC.
\bibitem{Z03} Zuo, Y. (2003) ``Projection-based depth functions and associated medians'',
\emph{Ann. Statist.}, 31, 1460-1490.
 \bibitem {Z04} Zuo, Y. (2004), ``Robustness of weighted $L_p$ -depth and $L_p$ - median'', {\it Allgemeines Statistisches Archiv}(Journal of the German Statistical Society), 88(1): 1-20.
\bibitem{Z10} Zuo, Y.\ (2010),
``Is $t$ procedure $\bar{x} \pm t_\alpha(n-1) s/\sqrt{n}$ optimal?",
\emph{The American Statistician},  64(2), 170-173.
\bibitem{Z09}  Zuo, Y. (2009), ``Data depth trimming procedure outperforms the  classical $t$ (or $T^2$) one",
\emph{Journal of Probability and Statistics}, http://dx.doi.org/10.1155/2009/373572.
\bibitem{Z18b} Zuo, Y. (2018), ``Robustness of deepest projection regression depth functional", {\it Statistical Papers}, https://doi.org/10.1007/s00362-019-01129-4,  arXiv:1806.09611.
\bibitem{Z18b} Zuo, Y. (2019a) ``Asymptotics for the maximum regression depth estimator", arXiv:1809.09896.
\bibitem{Z18b} Zuo, Y. (2019b), ``Computation of projection regression depth and its induced median", 	arXiv:1905.11846.
 \bibitem{ZS00a} Zuo, Y., Serfling, R., (2000), ``General notions of statistical depth function".  \emph{Ann. Statist.}, 28, 461-482.
\end{thebibliography}
\end{document}